\documentclass[useAMS,usenatbib]{mn2e}
\usepackage{epsfig,graphics}
\usepackage{color}

\def\rhoDM{\mbox{$\langle \rho_{\rm DM} \rangle$}}
\def\rhostar{\mbox{$\langle \rho_{\rm \star} \rangle$}}

\def\eSF{\mbox{$\epsilon_{\rm SF}$}}
\def\Re{\mbox{$R_{\rm eff}$}}

\def\Msun{\mbox{$M_\odot$}}

\def\ML{\mbox{$M/L$}}

\def\Ydyn{\mbox{$\Upsilon_{\rm dyn}$}}

\def\Mdyn{\mbox{$M_{\rm dyn}$}}
\def\mst{\mbox{$M_{\star}$}}

\def\fdm{\mbox{$f_{\rm DM}$}}
\def\lsim{\mathrel{\rlap{\lower3.5pt\hbox{\hskip0.5pt$\sim$}}
    \raise0.5pt\hbox{$<$}}}                
\def\gsim{~\rlap{$>$}{\lower 1.0ex\hbox{$\sim$}}}
\def\Msun{\mbox{$M_\odot$}}

\def\lsim{\mathrel{\rlap{\lower3.5pt\hbox{\hskip0.5pt$\sim$}}
    \raise0.5pt\hbox{$<$}}}
\def\gsim{~\rlap{$>$}{\lower 1.0ex\hbox{$\sim$}}}

\def\Re{\mbox{$R_{\rm eff}$}}

\def\eSF{\mbox{$\epsilon_{\rm SF}$}}

\def\SIS{\mbox{${\tt SIS}$}}
\def\trSIS{\mbox{{\tt tr-}${\tt SIS}$}}
\def\cMtoL{\mbox{{\tt const-}${\tt M/L}$}}

\def\Ymed{\mbox{$\langle Y \rangle$}}

\newcommand{\sn}{$n$}
\newcommand{\mie}{$<\! \mu\! >_{\rm e}$}

\def\mr{\mbox{$^{0.1}M_{r}$}}

\title[SPIDER dark matter]{SPIDER - VI. The Central Dark Matter Content of Luminous Early-Type
Galaxies: Benchmark Correlations with Mass, Structural Parameters and Environment}
\author[Tortora et al.]{\noindent
C.~Tortora$^{1}$\thanks{E-mail: ctortora@physik.uzh.ch}, F.~La
Barbera$^{2}$, N.R. Napolitano$^{2}$, R.R. de Carvalho$^{3}$ \and
A.J.~Romanowsky$^{4}$
\\~\\
$^1$ Universit$\ddot{a}$t Z$\ddot{u}$rich, Institut f$\ddot{u}$r
Theoretische Physik, Winterthurerstrasse 190, CH-8057,
Z$\ddot{u}$rich, Switzerland\\
$^2$ INAF -- Osservatorio Astronomico di Capodimonte, Salita
Moiariello 16, I-80131 - Napoli, Italy\\
$^3$ Instituto Nacional de Pesquisas Espaciais/MCT, S. J. dos
Campos, Brazil\\
$^4$ University of California Observatories, Santa Cruz,
CA 95064, USA \\}
\begin{document}
\date{Accepted  Received }
\pagerange{\pageref{firstpage}--\pageref{lastpage}} \pubyear{xxxx}
\maketitle
\label{firstpage}
\begin{abstract}
We  analyze  the central  dark-matter  (DM)  content  of $\sim
4,500$ massive ($M_\star \gsim  10^{10} \, M_\odot$), low-redshift
($z<0.1$), early-type galaxies  (ETGs), with high-quality
$ugrizYJHK$ photometry and  optical  spectroscopy from  SDSS  and
UKIDSS.   We estimate  the ``central'' fraction of DM  within the
$K$-band effective radius, \Re, using spherically symmetric
isotropic galaxy models.  We discuss  the  role  of  systematics
in  stellar  mass  estimates,  dynamical modelling, and velocity
dispersion  anisotropy. The main results of the  present  work
are  the  following:  (1)  DM  fractions  increase systematically
with both structural parameters (i.e. \Re, and S\'ersic index,
$n$) and mass proxies (central velocity dispersion, stellar and
dynamical  mass), as in  previous studies,  and decrease  with
central stellar  density.   2) All  correlations  involving  DM
fractions  are caused  by  two fundamental  ones  with  galaxy
effective radius  and central  velocity dispersion.  These
correlations are  independent of each other,  so that  ETGs
populate a  central-DM plane (DMP),  i.e. a correlation among
fraction of total-to-stellar mass, effective radius, and velocity
dispersion, whose scatter along the total-to-stellar mass axis
amounts to  $\sim 0.15$~dex.   (3)  In general,  under  the
assumption of an isothermal or a constant \ML\ profile for the
total mass distribution,  a Chabrier  IMF is favoured  with
respect  to a bottom-heavier   Salpeter  IMF,  as   the  latter
produces  negative (i.e. unphysical) DM fractions for more than
$50\%$ of the galaxies in our  sample.   For  a  Chabrier  IMF,
the  DM  estimates  agree  with $\Lambda$CDM  toy-galaxy models
based on  contracted  DM-halo density profiles. We also find
agreement with predictions from hydrodynamical simulations.   (4)
The  central DM  content  of ETGs  does not  depend significantly
on the environment where galaxies reside, with group and field
ETGs having similar DM trends.
\end{abstract}
\section{Introduction}\label{sec:intro}
In the  last decade, large area  surveys like SDSS  (Sloan Digital Sky
Survey;  \citealt{DR1,DR6,DR7}),  have   provided  high  quality  data
contributing significantly  to our understanding  of galaxy properties
and scaling  relations. Nevertheless, galaxy formation  remains one of
the outstanding questions of modern astrophysics.  As the most massive
stellar  systems in  the nearby  universe, early-type  galaxies (ETGs)
have a  special role in  providing the underpinnings for  a consistent
galaxy formation picture. They  form a relatively homogeneous class of
objects, dominated by  an old stellar population, a  small fraction of
cold  gas  and  low-levels  of  ongoing star  formation  (SF).   These
characteristics  make them a  potentially powerful  tool to  trace the
evolution of  cosmic structures back  through the cosmic  epochs.  The
uniformity  in  the  ETG  properties involves  tight  correlations  of
quantities  like effective  radius, \Re,  surface  brightness measured
within this radius, and central velocity dispersion, $\sigma_0$, which
merge  into  the   so-called  Fundamental  Plane  (FP;  \citealt{DD87,
  Dressler87}).  An observed deviation between the FP coefficients and
those  expected from the  virial theorem  has been  interpreted, among
other possibilities,  as a variation of  the total \ML  \, with galaxy
luminosity/mass (\citealt{Dressler87})  which in turn  may reflect the
DM    content    of     an    ETG    (see    e.g.     \citealt{CLR96};
\citealt{Busarello+97};         \citealt{GC97};        \citealt{PS97};
\citealt{TBB2004,      Cappellari06,      DOnofrio+06,      Graves09};
\citealt[T+09]{Tortora2009a};   \citealt{LaBarbera+10b}).   Therefore,
measuring  the  DM content  in  an  independent  way is  of  paramount
importance  for studying  the ETG  scaling  relations as  well as  the
overall process  of galaxy formation and evolution.   Recently, the DM
content in the central regions  of ETGs (typically within $1 \Re$) has
been  analyzed  using  both  local  samples  (\citealt{Padmanabhan04};
\citealt{Cappellari06};   \citealt{HB09};   T+09;   \citealt[hereafter
  NRT10]{NRT10}) and intermediate-redshift gravitational lens galaxies
(\citealt{Cardone+09};        \citealt{SLACSX};        \citealt{CT10};
\citealt[hereafter    T+10]{Tortora+10lensing};    \citealt{Faure+11};
\citealt{More+11}). Independent of the model used to describe the mass
distribution in a galaxy, several studies have found that DM fractions
within \Re\  increase with galaxy luminosity, stellar  mass, size, and
velocity  dispersion (e.g.,  \citealt{FSW:05}; \citealt{Cappellari06};
\citealt{Forber+08};        T+09;       NRT10;       \citealt{SLACSX};
\citealt{Leier:11};  T+10),  while a  different  conclusion was  drawn
by~\citet{TBB2004},  based on  a  constant-\ML\ mass  model (see  also
\citealt{2010ApJ...722..779G}).   The correlation  of DM  density with
stellar  mass   and  \Re  \,   points  to  DM  profiles   being  cuspy
(\citealt{Thomas+09}; T+09; NRT10; T+10).  However, even in this case,
inconsistent findings  have been  reported.  Using rotation  curves of
spiral  galaxies,  and mass  models  of  individual  dwarf and  spiral
galaxies,  as well  as  the  weak lensing  signal  of ellipticals  and
spirals,  \cite{Donato+09} and  \cite{G09} found  that the  central DM
column  density  is  constant  over  twelve  orders  of  magnitude  in
luminosity.   On the  contrary,  NRT10 showed  that,  on average,  the
projected central density of nearby ETGs is systematically higher than
that of spiral and dwarf  galaxies, implying an increase of DM density
with halo mass.  The same conclusion was reached  by \cite{B09}, using
data for different galaxy types and groups/clusters of galaxies.  Part
of  the above controversies  may be  associated with  sample selection
issues, with differences in the way galaxy parameters are measured, as
well as  with different  assumptions about the  halo models,  IMF, and
adopted fiducial radius~\citep{CT10}. In  this paper, we study how the
central DM  content of ETGs  correlates with galaxy properties,  for a
large  and  homogeneous  sample  with  a  wealth  of  photometric  and
spectroscopic    data     available    \citep[Paper    I     of    the
  series]{LaBarbera+10a}.  In  previous  papers,  we analyzed  the  FP
relation   of  ETGs   and   its  dependence   on  galaxy   environment
(\citealt[Paper           II]{LaBarbera+10b};           \citealt[Paper
  III]{LaBarbera+10c});  and   the  correlation  of   internal  colour
gradients   of    ETGs   with   galaxy    properties   (\citealt[Paper
  IV]{LaBarbera+10b}).  Here, we investigate the DM content of ETGs in
terms  of  both  structural   parameters  and  various  mass  proxies,
contrasting  the average  trends  with predictions  of toy-models  and
cosmological  simulations.   We show,  for  the  first  time, how  the
correlations between  the DM content  and galaxy properties  depend on
the  environment  where  these  systems reside.  Our  study  resembles
previous ones in its use of SDSS data to study the connections between
stellar                and                dynamical               mass
\citep{Padmanabhan04,2009MNRAS.396L..76S,GF10,2010ApJ...722..779G},
but goes beyond these  by incorporating additional photometric data as
well  as   environmental  information,  and  may   be  considered  the
definitive study of the central DM content of a large sample of bright
ETGs in  the SDSS  era.  The outline  of the  paper is as  follows. In
Section \ref{sec:sample} we describe  the sample and the data analysis
as  well  as stellar  and  dynamical  mass  calculations.  In  Section
\ref{sec:DM},  central  DM fraction  and  density  are  analyzed as  a
function  of  galaxy  mass   and  structural  parameters.  In  Section
\ref{sec:DM_plane} we discuss  the main drivers of the  DM content and
the  ``DM  plane''  of  ETGs,  and in  Section  \ref{sec:FP_tilt}  the
implications for the FP are discussed.  Section \ref{sec:DM_env} deals
with  galaxy environment,  while Section  \ref{sec:DM_models} compares
the   correlations  involving  DM   content  with   expectations  from
cosmological  and  toy-galaxy  models.  Conclusions are  presented  in
Section \ref{sec:conclusions}. Throughout the paper, we adopt $H_{0} =
75 \, \textrm{km} \, \textrm{s}^{-1} \, \textrm{Mpc}^{-1}$.
\section{Data sample and analysis}\label{sec:sample}
\subsection{Sample}
A volume-limited sample  of $39, 993$ ``bright''
(${}^{0.1}M_r<-20$) ETGs, in  the  redshift  range  of $0.05$  to
$0.095$,  with available  $ugriz$ photometry  and  optical
spectroscopy from SDSS-DR6 is used in this work and in previous
papers of the SPIDER (Spheroid's Panchromatic Investigation in
Different Environmental Regimes) project.  $5,080$ galaxies  also
have $YJHK$ photometry from DR2 of  UKIDSS-LAS (see  Paper I).
ETGs are  defined as bulge dominated systems, with passive
spectra in their centres (i.e.   within the SDSS fibre apertures).
Following \cite{BER03a}, from an operational viewpoint, ETGs are
those systems with {\tt eClass}  $<  0$ and  {\tt  fracDevr}  $>
0.8$, where  the  SDSS spectroscopic  parameter {\tt  eClass}
gives  the spectral  type of a galaxy, while  the SDSS photometric
parameter {\tt fracDevr} measures the fraction of galaxy light
that is better fitted by a de Vaucouleurs (rather than an
exponential) law\footnote{{Notice that the {\tt eClass} and {\tt
fracDevr} selections are very effective to remove late-type
systems (see Paper I), but do not allow a clear separation of E
and S0 galaxy types. Since S0s are flatter than ellipticals, we
have carried out a test of restricting the analysis to objects
with $K$-band axis-ratio $q > 0.8$. With this alternative cut that
minimizes the fraction of S0s, we find no significant variation in
the results presented throughout the paper.}}. All galaxies have
central velocity dispersion, $\sigma$, from SDSS-DR6, in the range
of $70$ to $420$~km~s$^{-1}$.   In all wavebands, galaxy
structural parameters -- i.e.  the effective radius, \Re, the mean
surface brightness within that  radius, \mie,  and  the S\'ersic
index, \sn\  -- have been homogeneously measured using
2DPHOT~\citep{LBdC08}, fitting galaxy images with seeing convolved
two-dimensional S\'ersic models.   The SPIDER sample is $95\%$
complete at \mr$=-20.32$, or, at  a stellar mass $M_\star = 3
\times 10^{10} M_\odot$ for a \citet{Chabrier01} IMF.

For  the present  study, we  select  SPIDER ETGs  with high
quality structural  parameters in the optical  and Near-InfraRed
(NIR) wavebands,  according to  the  following criteria: the
S\'ersic fit has $\chi^2<2$ in all wavebands; uncertainty on $\log
\Re$ $<$ $0.5$~dex from $g$ through $K$ (see Paper IV for details
on these thresholds); available stellar mass estimates
(\citealt[Paper V]{Swindle+11}). The resulting sample consists of
$4,259$ ETGs.
As expected,  all these  galaxies reside on the red-sequence, with
more than $99 \%$ having $g-r \gsim 0.5$ (within an aperture of
1~\Re), and a  median $g-r=0.88$.  For this  sample, the \Re\
spans the range $\sim (0.5 - 40) \, \rm kpc$ ( $\sim (1 - 15)\,
\rm kpc$ for more than $90\%$ of  the galaxies), with a  median of
$3.5 \,  \rm kpc$. The median ratio between  the SDSS fibre
aperture, where  velocity dispersions  are measured,  and the
$K$-band effective  radius, $R_{\rm ap}/\Re$, amounts  to $\sim
0.6$, implying only  a little ``extrapolation''    in   our DM
fraction   estimates (see Sec.~\ref{sec:Dmasses}).
As shown in
Paper I, significant differences are found when comparing 2DPHOT
and SDSS structural parameters.  Such differences arise from the
use of S\'ersic (2DPHOT) rather  than de Vaucouleurs (SDSS) models
to  fit the light and total mass distribution of  ETGs; from
different software  to measure the structural parameters;  and
from the sky estimation bias that affects SDSS
photometry~\citep{DR6}.  In Appendix A,  we further illustrate
this  point  by comparing correlations  among structural
parameters from different sources.
The environment  of ETGs in the  SPIDER sample is  characterized
by a friends-of-friends catalog of $8,083$ groups,  created as in \citet{Berlind:06}.
Here, we used  Berlind's algorithm over a larger
area, as the new SDSS-DR7 is now available (rather  than DR3; see
Paper  III  for details). A shifting  gapper technique is applied
to this catalogue (see~\citealt{lop09a}), allowing galaxies to  be
classified  as either  group  members  ($\sim46\%$), {non-group
members (hereafter ``field'' galaxies}; $\sim  33 \%$), or
unclassified ($\sim 21 \%$).  We also separate group members into
central  and satellite galaxies, where the central galaxy of a
given group is the object with the largest stellar
mass~\citep{Yang:07}.
\subsection{Stellar mass estimates}
\label{sec:Smasses}
Stellar masses are derived  by  fitting  synthetic stellar
population (SP) models from~\citet[BC03 hereafter]{BC03} to the
optical+NIR photometry, using the software {\it LePhare}
(\citealt{Ilbert+06}).  The observed galaxy fluxes  determine the
normalization of  the best-fit  template, which then gives the
stellar mass, \mst\ (initially computed within the SDSS Kron
aperture, which we correct using S\'ersic models to total mass).
We adopt a set  of SP models with different star-formation
$e$-folding times  ($  \tau  \le  15$~Gyr), internal reddenings
($E(B-V) \le  0.5$), and metallicities  ($0.2 \le Z/Z_{\odot} \le
3$), assuming a \citet{CCM89} extinction law and Chabrier IMF.

\begin{figure*}
\psfig{file=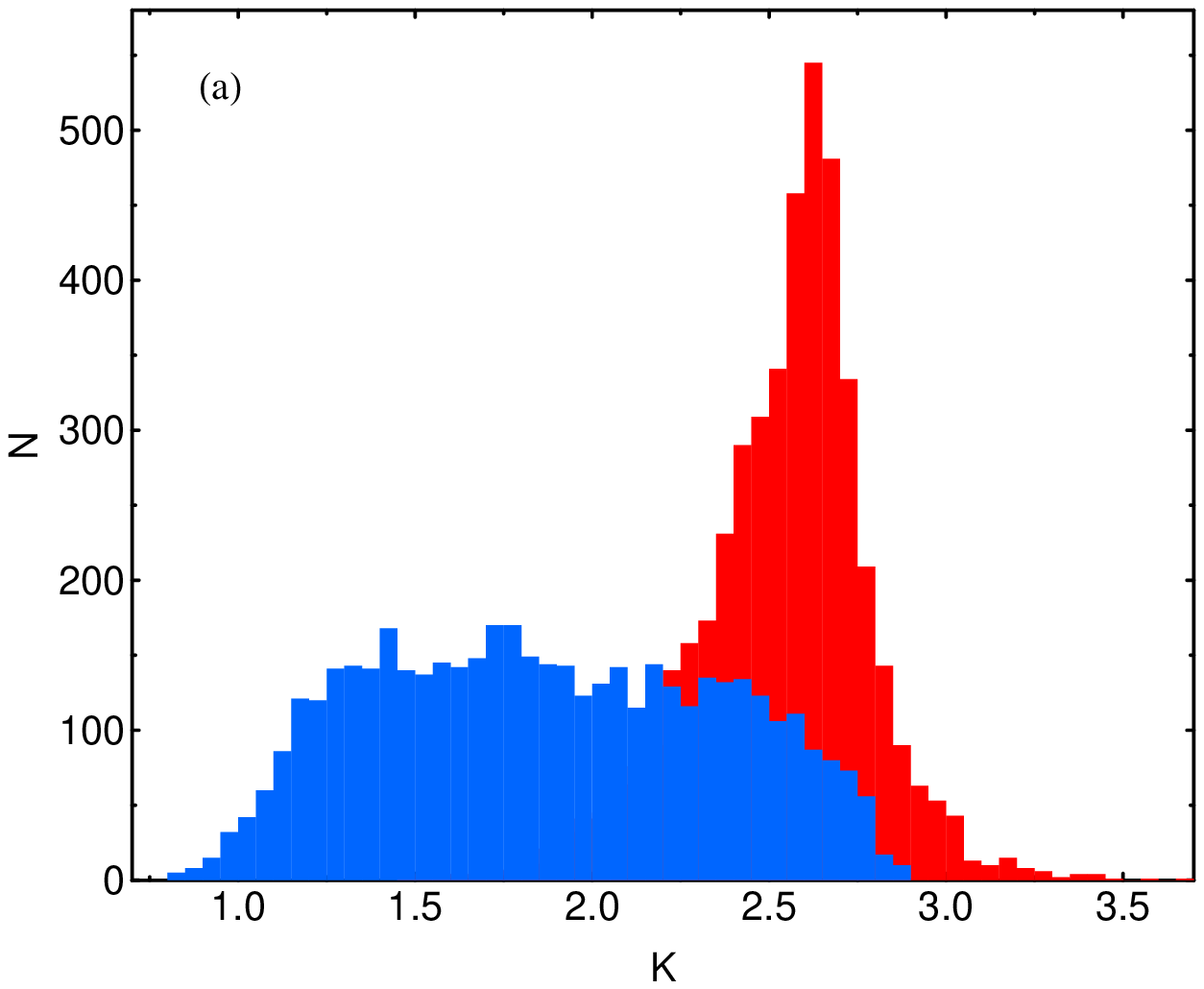, width=0.53\textwidth}\psfig{file=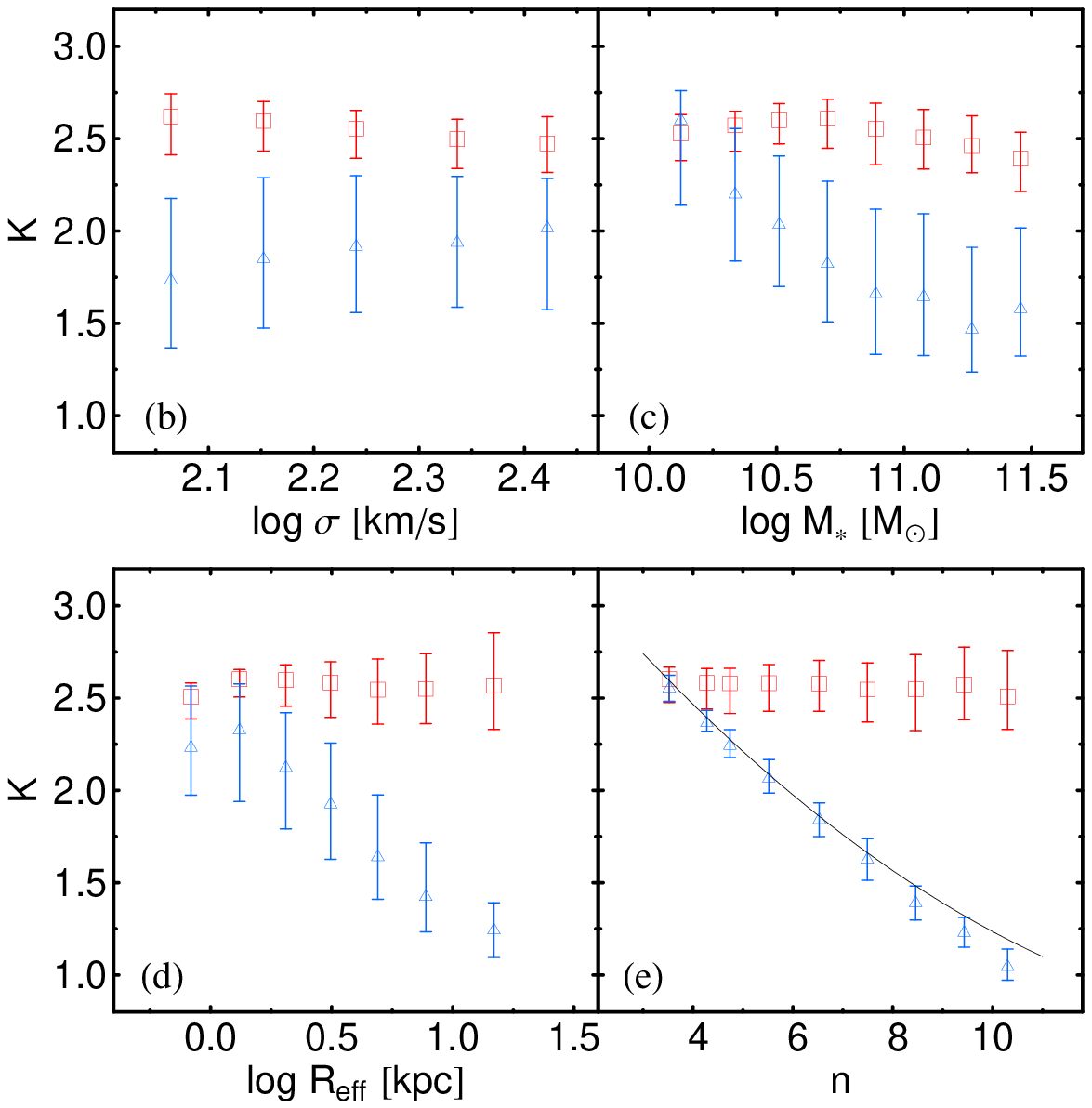,
  width=0.46\textwidth}\caption{{\it Panel (a).}  Distributions of $K$
  for \SIS\ and \cMtoL\ mass models (see Eq.~(\ref{eq:M_virial})).  {\it
    Panels   (b-e).}   Median   value  of   $K$  as   a   function  of
  $\sigma$ (b), \mst\ (c),  \Re\ (d), and $n$ (e).  Median
  values, with error bars showing 25--75 per cent scatter.   Red and blue
  colours refer  to \SIS\ and \cMtoL\ models,  respectively.  The black
  curve in panel (e) is taken from \citet{BCD02}.}\label{fig:figK}
\end{figure*}

We  refer the  reader to  Paper V  for all  details about  the
fitting procedure and  the accuracy of stellar mass  estimates
using different theoretical and  empirical assumptions.  We
provide here  only a
  summary of statistical and systematic uncertainties on stellar mass
  estimates.   In paper  V, we  compared  stellar  mass estimates
  obtained with  different stellar  population models, e.g.   BC03 and
  PEGASE.2  models~\citep{FRV97}, different internal  extinction laws,
  and different  combinations of wavebands  adopted in the  SED fitting
  procedure.    Also,    we performed comparisons with
  spectroscopically derived stellar masses  (the two sets of estimates
  being in  excellent agreement, with  a median difference  of $0.03\,
  \rm dex$).

In  general, the comparison of different  sets of stellar
  mass  estimates   shows  that  the  scatter  in   \mst\  (i.e.   its
  uncertainty) ranges  from $0.05$ to $0.15$~dex.
  As in other studies (see Paper V and references therein)
  systematic uncertainties on stellar mass  were found to play a
  major role, with  variations in the IMF and  extinction law yielding
  systematic biases on the mass of nearly a factor of 2.  Notice though,
  that despite  the  age-metallicity degeneracies  in  photometric data,
  these conspire  to keep the  stellar mass-to-light ratio  (and hence
  the stellar masses) relatively well constrained.
A de-projected S\'ersic law in the $K$-band is used to describe
the density profile of the stellar component.

\subsection{Dynamical mass estimates}
\label{sec:Dmasses}

The dynamical mass within a given galaxy radius, $r$, is usually computed as:
\begin{equation}
M_{dyn}= \frac{K \sigma^{2} r}{G},\label{eq:M_virial}
\end{equation}
where  $G$  is the  gravitational  constant,  $\sigma$  is the  galaxy
velocity  dispersion, and  $K$  is a  pressure  correction term  (e.g.
\citealt{Padmanabhan04, Eke04, Cappellari06};  T+09).  The $K$ depends
on several factors, like the  radius wherein \Mdyn \, is computed, the
aperture used  to measure $\sigma$,  the viewing angle of  the system,
its   orbital   structure,  luminosity   profile,   and   how  DM   is
distributed. Here, we follow the  approach used in T+09, where instead
of adopting some approximation  for $K$ in Eq.~(\ref{eq:M_virial}), we
model  each individual galaxy  directly using  the Jeans  equations to
estimate  \Mdyn\ within $r=$~1~\Re\  (see also  \citealt{CT10}; T+10).
The models also require assumptions about the mass profile in order to
extrapolate  to  \Re\ from  the  more  central  (on average)  $\sigma$
measurements. Here our approach is  to adopt two types of mass profile
that bracket  a range  of possibilities. Each  of these  describes the
total mass profile with one free parameter:
\begin{itemize}
\item {\it  SIS.} The  Singular Isothermal Sphere,  where $M(r)\propto
  \sigma_{\rm  SIS}^{2}  r$, is  the  prototype  of  a galaxy  profile
  producing  a  flat  rotation  curve. Despite  its  simplicity,  this
  reproduces quite  well the  total mass profile  in the  most massive
  ETGs (e.g.  \citealt{Kochanek91, SLACS3}; \citealt{SLACSIV}), and in
  particular  the  massive  halo   present  in  their  outer  regions,
  consistently with  virial mass estimates (\citealt{Benson2000,MH02};
  \citealt{Nap+05};  \citealt{vdeB07}).    Since  a  galaxy   halo  is
  expected to be truncated when it enters a group/cluster of galaxies,
  we analyze the impact of the environment on the central DM fractions
  using also a truncated SIS model (\trSIS), where the density profile
  is truncated at  a radius $r_{t}$.  For each of  the group ETGs, the
  value of $r_t$  is computed from the projected  distance between the
  ETG and  the centre  of the parent  group, following the  recipes of
  \citet{Ghigna+98}.
\item {\it Constant \ML.}  Some  studies of ordinary ETGs suggest that
  the mass profile falls off roughly as steeply as the stars, at small
  to intermediate  radii, which  may imply either  little DM  in their
  centres     or    DM     profiles    that     mimic     the    stars
  \citep{R+03,Douglas+07,Deason12}. The mass profile is given by $M(r)
  =   \Upsilon_0  \,   L(r)$,  where   $\Upsilon_0$  is   the  typical
  mass-to-light  ratio of  the  system and  $L(r)$  is the  luminosity
  profile, modeled  here by the \cite{Sersic68} law.   Hence, the mass
  distribution  is  assumed  to  follow  the profile  of  the  stellar
  component, whose shape (i.e.  the S\'ersic $n$), changes from galaxy
  to  galaxy,  reflecting the  structural  non-homology  of the  light
  profiles of ETGs.
\end{itemize}

Operationally, for each galaxy, given a mass profile model, we compute
the  projected   mass-weighted  velocity  dispersion   of  the  model,
$\sigma_{model}$,    within   the    SDSS    fibre   aperture.     The
$\sigma_{model}$  is then  matched  to the  observed central  velocity
dispersion  of  the   galaxy,  providing  the  corresponding  best-fit
parameter,  $\sigma_{\rm SIS}$ ($\Upsilon_0$)  for the  \SIS\ (\cMtoL)
model,   and   the   mass    profile   $\Mdyn   (r)$   (see   Appendix
\ref{sec:Jeans_procedure} for details).

\section{Dynamical masses and central dark matter}\label{sec:DM}
We characterize the DM content of an ETG  by  computing (i) the
de-projected DM  fraction, $\fdm(r) =  1 - \mst(r) / \Mdyn(r)$ or
the total-to-stellar mass ratio $\Mdyn(r)/\mst(r)$, where \mst \,
and \Mdyn \, are the stellar and dynamical mass~\footnote{We
implicitly assume that the dust and gas components give a
negligible contribution to \Mdyn.} within a given de-projected
radius, $r$; and (ii) the de-projected average DM density, $\rhoDM
= ( \Mdyn(r)  - \mst(r) ) / (\frac{4}{3}\pi r^{3})$,  giving more
direct information on the DM content of a galaxy. The $\mst(r)$ is
estimated by de-projecting the $K$-band light profile of each
galaxy (assuming spherical symmetry), and normalizing this
de-projected profile to the total \mst \, estimate, from Paper V.
Hereafter,  we  refer to $\fdm(r=R_e)$ ($\langle \rho_{DM}(r=R_e)
\rangle$), i.e. the DM fraction (density) computed within a
de-projected radius equal to the projected $R_e$, as the
``central'' DM fraction (density).
\subsection{Dynamical masses and virial coefficients}
\label{sec:K}
To provide a more straightforward way of estimating \Mdyn\ from
velocity dispersion and effective radius (as done in previous
work), and of comparing mass estimates from different models, we
also recast the computation   of   dynamical masses in terms of
Eq.~(\ref{eq:M_virial}). For  each  galaxy, we  correct  the
observed $\sigma$  to  an aperture  of 1~\Re, $\sigma(\Re)$,
following~\citet{Cappellari06} (see also
\citealt{Jorgensen+95,Jorgensen+96}).  Then, we insert \Re,
$\sigma(\Re)$, and the dynamical mass estimates obtained through
the Jeans equations (see above) into Eq.~(\ref{eq:M_virial}) to
obtain the corresponding virial  coefficients, $K_{\rm SIS}$ and
$K_{\rm M/L}$, for the \SIS\ and \cMtoL \, models, respectively.
The distributions of $K_{\rm SIS}$ and $K_{\rm M/L}$  are
displayed in the left--panel of Fig.~\ref{fig:figK}, while the
right panels show the median values of $K_{\rm  SIS}$ and  $K_{\rm
M/L}$  as a function of $\sigma$, \mst, \Re, and  $n$. The
distributions of $K_{\rm SIS}$ and $K_{\rm M/L}$ have median
values~\footnote{Using an aperture of  $\Re/8$, rather  than
$1\Re$,  gives $K_{\rm SIS} = 2.14^{+0.11}_{-0.18}$ and $K_{\rm
M/L}=1.54^{+0.38}_{-0.35}$.} of $K_{\rm SIS}
=2.57^{+0.11}_{-0.17}$  and $K_{\rm M/L}=1.85^{+0.43}_{-0.38}$
(where error bars  are the 25--75 per cent scatter). The $K_{\rm
M/L}$ is always  smaller  than  $K_{\rm SIS}$, reaching equality
only for galaxies with a $K$-band S\'ersic $n \sim 3.5$ (see panel
(e) of Fig.~\ref{fig:figK}). This is consistent with the findings
in T+09, where galaxy  structural parameters were derived  using a
de Vaucouleurs rather than a S\'ersic model.
From Fig.~\ref{fig:figK} we see that $K_{SIS}$ is independent  of
$\sigma$, \mst, \Re, and $n$, while $K_{\rm M/L}$ decreases with
\mst, \Re, and $n$, and is independent of $\sigma$. The $K-n$
trend is in fairly good agreement with the best-fitting $K-n$
relation obtained by \cite{BCD02} (see solid black line in panel
(e)).  Notice  that using a constant $K_{SIS}$ rather than fitting
each  individual galaxy with the \SIS\  model would introduce an
uncertainty in dynamical mass equal to the scatter of the
$K_{SIS}$ distribution in  Fig.~\ref{fig:figK}, $\sim4-7\%$
(taking the lower and upper 25--75 per cent range, respectively).
Similarly, using the $K-n$ median trend  to obtain  $K_{\rm M/L}$
would imply an uncertainty of $\sim  4$--$5\%$ on \Mdyn. Modelling
each individual galaxy has the advantage of avoiding this source
of scatter, which could be important for discerning subtle
differences in DM content among different subsamples (e.g.
different environments, see Sec.~\ref{sec:DM_env}).

For group ETGs, we have also computed $K$ values by using the
truncated \SIS\ model,  \trSIS\ (see Sec.~\ref{sec:Dmasses}),
finding them to be fully consistent with those obtained from the
\SIS\ model.  In particular,  only $\lsim 2 \%$ of  group galaxies
exhibit a difference of  $\Delta \log  K  > 0.01$ ($\sim 2.3 \%)$.
This is due  to the fact that the truncation radius, $r_t$ (see
Sec.~\ref{sec:Dmasses}), has a distribution peaked at  $\sim
50$~kpc, with \Re\ being $\sim 3.5 $~kpc, i.e. $\ll r_{t}$.

Throughout the present paper,  the DM content of ETGs is
  estimated  under  the  assumption  that  their stellar orbits are
  isotropic, which is incorrect at some level.  Although a
  detailed analysis  of anisotropic orbits is far  from being trivial,
  and  is certainly  beyond the  scope of  the present  work,  we have
  estimated how  anisotropy can  affect our dynamical  mass estimates.

  Detailed dynamical modeling efforts of ETGs have focused extensively
  on their central regions
  (i.e.   those   we  investigate   in  this  work), and found
  anisotropies to be fairly mild in general, typically
   in the range $-0.2 \le \beta \le +0.3$ \citep{Gerhard+01,Cappellari07},
  where $\beta \equiv 1 - \sigma_\theta^2/\sigma_r^2$ quantifies
  the relative internal dispersions in the tangential and radial directions.
  Mild central anisotropy is also predicted from simulations of
  merger remnants \citep{Dekel+05}.

  Recent work has focused attention on the ability to constrain
  $\Mdyn(\Re)$ in stellar systems, independently of anisotropy,
   {\it if the velocity dispersion can be measured over the whole galaxy}
   \citep{Walker09,Wolf+10}.
  If the dispersion is measured only within an aperture of \Re, but
  the dispersion profile is relatively constant with radius, it is
  possible that the mass can still be fairly well constrained
  (G. Mamon, priv. comm.).  To check this, we have carried out
  test models as in Appendix~B, but adopting constant anisotropy profiles,
  $\beta(r) = \beta$.

  For $\beta = 0.1$ ($=0.2$),
  we find  dynamical masses that  are $\sim 2\%$ ($\sim  4\%$) smaller
  than in  the isotropic case.  In  the case of  extremely (and indeed
  unrealistic) radial (tangential) orbits, i.e.  $\beta = 1$ ($\beta =
  -1$), the  masses are underestimated (overestimated)  by $\sim 30\%$
  ($\sim  10\%$).  The  effect  is  larger for  the  most massive  and
  largest galaxies in our sample, for which we find that $\beta = 0.1$
  ($=0.2$) would make  the masses smaller by $\sim  3\%$ ($\sim 6\%$),
  while in the case of $\beta = 1$ ($\beta = -1$), the masses would be
  smaller (larger) by $\sim  50\%$ ($\sim 20\%$).  Thus, for typically
  observed  values of  $\beta$ ($  \sim 0.2$),  we do  not  expect any
  significant variation  in the DM  content of ETGs as  inferred under
  the  assumption   of  isotropy, the  effect   being  $\lsim
  5\%$.

\subsection{Dark matter fractions}
\label{sec:DM_fraction}
\begin{figure}
\psfig{file=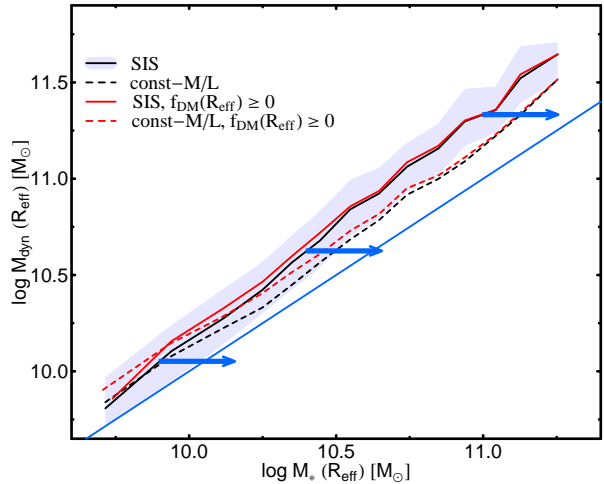, width=0.47\textwidth}\caption{ Dynamical
mass within 1~\Re\ as a function of stellar mass within the same
radius. The black solid (dashed) line is the median \SIS\ (\cMtoL)
trend, with the shaded region corresponding to the 25--75 per cent
scatter for the \SIS\ case. The red lines are the median trends
obtained by using only galaxies with $\fdm \geq 0$. A Chabrier IMF
is assumed in the computation of \mst\ (see Paper V). The solid
blue line is the one-to-one relation $\Mdyn(\Re) = \mst(\Re)$. The
arrows show the variation of \mst\ when adopting a Salpeter,
rather than a Chabrier IMF.} \label{fig:Mdyn_Mstar}
\end{figure}
Fig. \ref{fig:Mdyn_Mstar} plots median  dynamical masses as a function
of stellar mass, $\mst(\Re)$, for  both \SIS\ and \cMtoL \, models. We
find  $\Mdyn(\Re)  \propto  \mst(\Re)^{\alpha}$,  with $\alpha  >  1$,
i.e. the central DM content increases with galaxy mass. In particular,
we find $\alpha_{\rm  SIS} \sim 1.2$ and $\alpha_{\rm  M/L} \sim 1.07$
for  the \SIS\  and \cMtoL  \, models,  respectively.  Notice  that on
average,  at  fixed $\mst  (\Re)$,  the  $\Mdyn(\Re)$  is larger  than
$\mst(\Re)$,  and  the  fraction  of  ETGs  with  \mst$>$\Mdyn\  (i.e.
unphysical  DM fractions)  is $\sim  12\%$  ($24\%$) for  the \SIS  \,
(\cMtoL).   Using a Salpeter  (rather than  Chabrier) IMF  would shift
$\mst$ {\rm to  larger values by $\sim 0.25$~dex  (see the blue arrows
  in the Figure)}, making the median $\Mdyn$ only slightly larger than
\mst, and increasing  the fraction of objects with  \mst$>$\Mdyn
\, up to $\sim  55\%$ ($78\%$) for the  \SIS \, (\cMtoL)  model.

Taken at face value, this  result implies that the bottom-heavy
Salpeter IMF is disfavoured with respect  to a Chabrier IMF, in
agreement with previous findings from stellar dynamics (e.g.,
\citealt{Cappellari06}; NRT10), and from lensing \citep{FSB:08,
FER:10}. However, we  warn the reader  that (i) most  of the
objects with \mst$>$\Mdyn\ are from a region of parameter space
(low   \mst)  where our  sample  is   more  affected  by
incompleteness  (see below); and (ii)  our analysis  relies  on
the assumption of a  given model for the dynamical  mass
distribution in ETGs  (i.e., either \SIS\  or  \cMtoL).  Indeed,
if  we relax  the assumption of  a universal  IMF, our data  would
be  consistent with either  a  Salpeter or  Chabrier  IMF  at high
mass  (T+09,  NRT10, \citealt{Auger+10a};   \citealt{Treu+10}) and
with   a  Chabrier   IMF preferred       at       low       mass
(\citealt{Barnabe+11}; \citealt{Sonnenfeld+11};
\citealt{Brewer+12,Cappellari+12,Dutton+12};
  see also Section~\ref{sec:noDM} below).

\begin{figure*}
\psfig{file=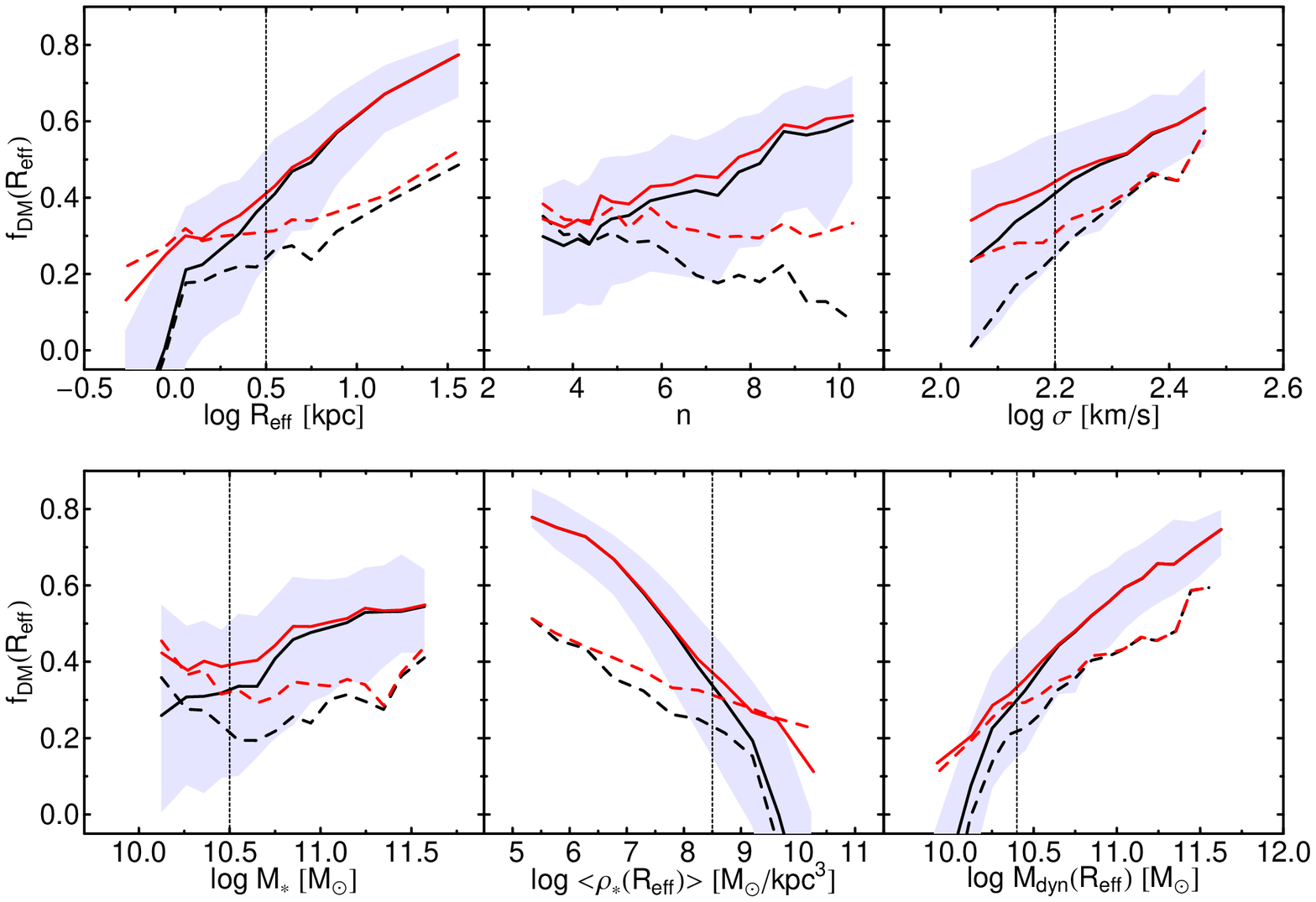, width=1\textwidth}\caption{DM fraction
within \Re\ as a function of $K$-band \Re, $n_{\rm K}$, velocity
dispersion $\sigma$, stellar mass \mst, central average stellar
density \rhostar\ and dynamical mass, \Mdyn. The symbols are as in
Fig. \ref{fig:Mdyn_Mstar}. The vertical dotted lines set the
completeness limit determined as discussed in the
text.}\label{fig:fig2}
\end{figure*}
Fig.~\ref{fig:fig2}  plots  central  DM  fractions as  a  function
of different galaxy  parameters, i.e.  effective  radius, S\'ersic
index, velocity dispersion,  stellar and dynamical mass,  and
central average de-projected  stellar  density,   \rhostar,
defined  as  $\rhostar  = \mst(\Re)/(\frac{4}{3}\pi  \Re^3)$.
Black  solid  (dashed) lines  show median trends for the \SIS\
(\cMtoL) model, while grey regions are the 25--75  per   cent
quantiles  (i.e.   the  scatter)   for  the
  \SIS\ model.  Notice  that, in general, the scatter  goes from $\sim
  0.2$ at low \fdm\ ($\sim 0.2$) up to $\sim 0.1$ (or smaller) at high
  \fdm\ ($\sim 0.8$).  This is likely caused by the statistical
  uncertainties  on galaxy  parameters,  such as,  in particular,  the
  \mst.  In fact, a  typical statistical uncertainty of $\sim 0.1$~dex
  on \mst\  (see Sec.~\ref{sec:Smasses}) propagates into  an error of
  $\sim  0.18$ ($\sim 0.05$)  on \fdm,  for \fdm$\sim  0.2$ (\fdm$\sim
  0.8$).

For each galaxy parameter, we derive the completeness limit
of  the  sample with  respect  to that  parameter  by  looking at  its
correlation with  the SDSS $r$-band Petrosian magnitude  (which is the
main parameter  used to select SDSS spectroscopic  targets; see Papers
II and IV  for further details on this  approach).  These completeness
limits~\footnote{In each  plot the sample  is complete at $\log  \Re /
  kpc \geq  0.5 \, \rm  kpc$, $\log \sigma  / (km/s) \geq  2.2$, $\log
  \mst/\Msun \geq 10.5$, $\log  \rhostar / (\Msun/kpc^3) \leq 8.5$ and
  $\log \Mdyn / \Msun \geq 10.4$.  Because of the large scatter in the
  diagram with S\'ersic $n$ vs.  SDSS Petrosian magnitude, we were not
  able to define a reliable completeness limit with respect to the $n$
  parameter.   However,  as  shown  in Fig.~\ref{fig:fig2}  (see  also
  Fig.~\ref{fig:fig3}),  the correlation  of DM  fractions (densities)
  with  $n$ are  approximately linear  over the  entire range  of this
  parameter, implying  that the  completeness limit is  unimportant to
  characterize  these correlations.}   are marked  by  vertical dotted
lines in Fig.~\ref{fig:fig2}.  Notice that because of the scatter
  around  the  correlations between  \fdm\  and  the different  galaxy
  parameters,  the range  of  \fdm, as  seen  by the  black curves  in
  Fig.~\ref{fig:fig2},  is different  for different  quantities, being
  larger for \Re, \rhostar\           and
  \Mdyn\  ($0\lsim \fdm \lsim  0.8$),  than for  \mst,
  $\sigma$ and $n$ ($0.2 \lsim \fdm \lsim 0.6$).
All median  trends in the  Figure are modeled by  linear
least-squares fits, adopting $\log  \fdm$ as the dependent
variable,  and using only the    portions    of   each    diagram
where    the   sample    is complete \footnote{We consider linear
relations of the form $Y =
    a + b  X$, where $b$ is the slope.  A  bootstrap method is applied
    to estimate $b$ and its  uncertainty. In practice, we bin the data
    with  respect to  the variable  $X$.  For each  bin, we randomly extract $50 \%$ of
    the points  in that bin, computing the  corresponding median value
    of $Y$.  We perform $1000$ iterations, each time computing the $b$
    from a  linear fit  of the  median values of  $Y$ vs.   the median
    values of  $X$ in all different  bins. The final value  of $b$ and
    its  uncertainty are  the median  and $\sigma$  of the  distributions of
    slope values  among all  iterations.   }.   In these  ranges, the
$\log \fdm$ trends  are reasonably well described by  the linear
fits. The  slopes  of  the  best-fitted  relations  are  reported
in  Table \ref{tab:tab1},  where it should  be kept  in mind  that
the  range of completeness  corresponds to  those galaxies  which
required  the most extrapolation from the fibre aperture to \Re,
so the \fdm\ results are the most  model dependent.   We also
tabulate  the slopes  obtained by using  $\log \Mdyn/\mst$, rather
than $\log  \fdm$, as  the dependent variable in the fits. This is
relevant to connect DM fractions to the origin  of scaling
relations of  ETGs  (Sec.~\ref{sec:FP_tilt}).   Notice also, that
due to scatter  in DM content at a given
  point  of the parameter  space, studying  \fdm\ is  not the  same as
  studying \Mdyn$/$\mst. In  fact, considering the non-linear relation
  between \fdm\  and $\Mdyn/\mst$, computing  the average $\Mdyn/\mst$
  and plugging  it into  the definition  of \fdm\ is  not the  same as
  computing directly the average \fdm.

\begin{table}
\centering \caption{Slopes of the correlation between $\fdm$,
$M_{\rm dyn}/\mst$ and \rhoDM\ vs \Re, $n$, $\sigma$, $\mst$,
\rhostar\ and \Mdyn, for the \SIS\ and \cMtoL\ models. We also
show in parenthesis the slopes when $M_{\rm dyn}
> \mst$. The fits are performed taking into
account the completeness limits.}\label{tab:tab1}
\begin{tabular}{lcc} \hline
 & \SIS\ & \cMtoL\  \\
\hline
 $\fdm-\Re$ & $0.26\pm 0.02$  & $0.27 \pm 0.05$ \\
                        & ($0.24\pm 0.02$)  & ($0.18 \pm 0.04$) \\
 $\fdm-n$ & $0.39 \pm 0.11$  & $-0.60 \pm 0.11$ \\
                        & ($0.33\pm 0.07$)  & ($-0.22 \pm 0.06$) \\
 $\fdm-\sigma$ & $0.71 \pm 0.11$  & $1.17 \pm 0.18$ \\
                        & ($0.62\pm 0.10$)  & ($0.84 \pm 0.15$) \\
 $\fdm-\mst$ & $0.24 \pm 0.04$  & $0.29 \pm 0.06$ \\
                        & ($0.15\pm 0.03$)  & ($0.07 \pm 0.05$) \\
 $\fdm-\rhostar$ & $-0.11 \pm 0.01$  & $-0.11 \pm 0.01$ \\
                         & ($-0.11\pm 0.01$)  & ($-0.07 \pm 0.01$) \\
 $\fdm-\Mdyn$ & $0.28 \pm 0.02$  & $0.33 \pm 0.03$ \\
                         & ($0.27\pm 0.01$)  & ($0.26 \pm 0.03$) \\
 \hline
 $M_{\rm dyn}/\mst-\Re$ & $0.40\pm 0.03$  & $0.16 \pm 0.04$ \\
                        & ($0.38\pm 0.03$)  & ($0.12 \pm 0.05$) \\
 $M_{\rm dyn}/\mst-n$ & $0.27 \pm 0.04$  & $-0.24 \pm 0.05$ \\
                        & ($0.25\pm 0.05$)  & ($-0.13 \pm 0.05$) \\
 $M_{\rm dyn}/\mst-\sigma$ & $0.78 \pm 0.13$  & $0.76 \pm 0.14$ \\
                        & ($0.69\pm 0.12$)  & ($0.61 \pm 0.13$) \\
 $M_{\rm dyn}/\mst-\mst$ & $0.19 \pm 0.03$  & $0.10 \pm 0.02$ \\
                        & ($0.13\pm 0.03$)  & ($0.04 \pm 0.02$) \\
 $M_{\rm dyn}/\mst-\rhostar$ & $-0.17 \pm 0.01$  & $-0.06 \pm 0.01$ \\
                         & ($-0.16\pm 0.01$)  & ($-0.05 \pm 0.01$) \\
 $M_{\rm dyn}/\mst-\Mdyn$ & $0.36 \pm 0.03$  & $0.23 \pm 0.03$ \\
                         & ($0.34\pm 0.02$)  & ($0.20 \pm 0.03$) \\
 \hline
 $\rhoDM-\Re$ & $-1.91 \pm 0.06$  & $-2.16 \pm 0.09$ \\
              & ($-1.96 \pm 0.06$)  & ($-2.30 \pm 0.09$) \\
 $\rhoDM-n$ & $-1.70 \pm 0.16$  & $-4.23 \pm 0.22$ \\
              & ($-1.77 \pm 0.14$)  & ($-3.25 \pm 0.12$) \\
 $\rhoDM-\sigma$ & $1.06 \pm 0.61$  & $1.88 \pm 0.74$ \\
              & ($0.77 \pm 0.43$)  & ($0.96 \pm 0.76$) \\
 $\rhoDM-\mst$ & $-0.91 \pm 0.11$  & $-0.78 \pm 0.18$ \\
              & ($-1.08 \pm 0.08$)  & ($-1.28 \pm 0.12$) \\
 $\rhoDM-\rhostar$ & $0.70 \pm 0.02$  & $0.83 \pm 0.03$ \\
              & ($0.71 \pm 0.02$)  & ($0.87 \pm 0.03$) \\
 $\rhoDM-\Mdyn$ & $-1.05 \pm 0.08$  & $-0.68 \pm 0.17$ \\
              & ($-1.12 \pm 0.08$)  & ($-0.93 \pm 0.16$) \\
 \hline
\end{tabular}
\end{table}
In agreement with NRT10 and T+10, we find a tight and positive
correlation between \SIS\  \fdm\, and \Re, which may  be
interpreted as a physical aperture effect, where  a larger \Re\
subtends  a larger portion  of  a galaxy DM  halo (see  also
\citealt{SLACSX}). A steep correlation also holds between DM
fraction and  S\'ersic $n$, namely, galaxies  with steeper light
profiles have  higher central  DM  fractions. We also find that,
independent of the adopted mass proxy (i.e. $\sigma$, \mst,
\Mdyn), more massive galaxies have  the largest DM content ($\fdm
\sim 0.6$), while less massive ones have \fdm$\sim 0.3$.

These findings are consistent with  other results in the
literature, based  on different samples and methodologies
(\citealt{Padmanabhan04}; \citealt{Cappellari06}; \citealt{HB09};
T+09; \citealt{CT10}; NRT10). An exception is the conceptually
related SDSS analysis by \citet{2010ApJ...722..779G}, who found a
fairly weak \fdm-\Re\ and \fdm-$\sigma$ correlations, and an {\it
anti}-correlation between \fdm\ and \mst. These differences may be
attributed to the use of a constant $K$, while our effective $K$
values varied from galaxy to galaxy depending on their individual
S\'ersic parameters.
We also note that \cite{2010ApJ...722..779G}
discussed projected rather than de-projected DM fractions. He converted the SDSS
velocity dispersion to a \Re/8 circular aperture and adopted an
expression for the dynamical mass similar to our Eq.
\ref{eq:M_virial} in his equation 4. The $K = \pi$ factor in his
equation 4 is equivalent to $K=2$ if the dynamical mass is
de-projected, and is slightly lower than the average value $\sim
2.15$ we have reported above. There is also a difference in using
$K$-band versus $r$-band effective radii (discussed below) and in
other details related to the sample selection (such as the redshift
range).
Fig.  \ref{fig:fig2} also  shows a  sharp anti-correlation  between DM
content  and  central average  stellar  density,  which  has not  been
reported in the literature so  far. Galaxies with denser stellar cores
have  lower DM fractions  (i.e. $\fdm  \sim 0$  at $\rho_\star  \sim 3
\times 10^{9} \,  \rm \Msun kpc^{-3}$), while \fdm\  values as high as
$\sim 0.8$ are found at  the lowest densities ($\rho_\star \sim 10^{6}
\, \rm  \Msun kpc^{-3}$).  This trend results  from the fact  that, on
average,  higher  stellar densities  correspond  to smaller  effective
radii,  implying a lower  \fdm. For  comparison with  previous studies
(see  Sec.~\ref{sec:Dmasses}), Fig.~\ref{fig:fig2} also  plots results
obtained with the \cMtoL\ model.   The \fdm\ values are still found to
increase  with \Re,  but with  shallower slope  relative to  the \SIS,
while  the trend with  the S\'ersic  $n$ is  inverted. The  trend with
$\sigma$ is also similar to that  obtained with the \SIS, but at fixed
$\sigma$  the  \fdm\ values  are  smaller  by  $\sim 0.1$--$0.2$.  The
\cMtoL\ model implies no correlation of \fdm\ with stellar mass, and a
shallower  trend (relative  to  the \SIS)  with  both central  stellar
density and \Mdyn.

\begin{figure}
\psfig{file=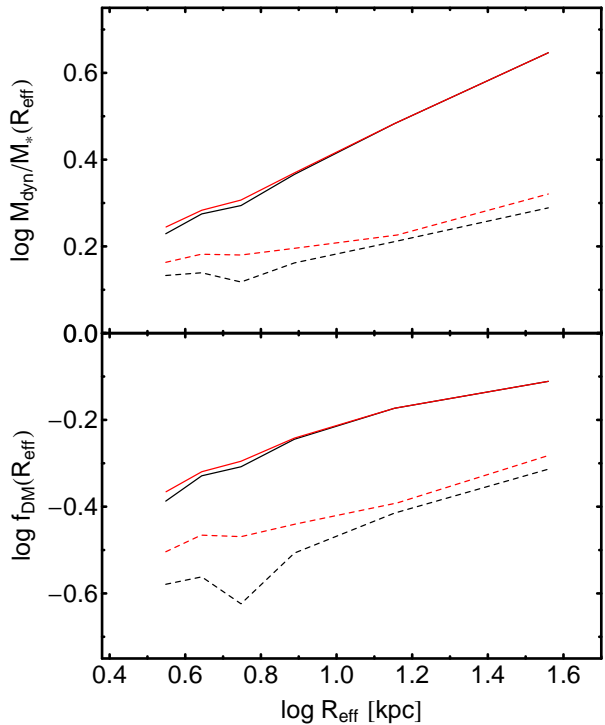, width=0.47\textwidth}\caption{ $M_{\rm
dyn}/\mst$ (top panel) and \fdm\ (bottom panel) within \Re\ in
terms of \Re\ for a \SIS\ and \cMtoL. The plot is limited to the
region where the sample is complete. The symbols are as in Fig.
\ref{fig:Mdyn_Mstar} and \ref{fig:fig2}. }
\label{fig:fig_fdm_MdMst}
\end{figure}

From the slope values reported in Table \ref{tab:tab1}, we notice
that using either  \fdm\ or $\Mdyn/\mst$  as the dependent
variable  in the fits leads to different conclusions  about the
comparison of \SIS\ and \cMtoL\ trends.  For  instance, the
correlation of \fdm\  with \Re\ is steeper  for \cMtoL\  than
\SIS,  while the  opposite  holds for  the $\Mdyn/\mst$--\Re \
correlation.  Whilst  the use of  \fdm\ and $\Mdyn/\mst$ is not
truly equivalent,  because of the scatter in DM content  at each
point  of  the parameter  space  (see above), the apparent
discrepancy seen  in Table  \ref{tab:tab1} for, e.g., the \fdm--
and $\Mdyn/\mst$--\Re\  relations, is indeed caused by the
non-linear relation between \fdm\  and $\Mdyn/\mst$: a difference
in \Mdyn/\mst\  (such  as  that between  \SIS\  and \cMtoL\
models) corresponds to a smaller difference  in \fdm\ at higher,
relative to lower, \Mdyn/\mst\ (i.e. going from low- to high-\Re\
values). See also Fig. \ref{fig:fig_fdm_MdMst} where this effect
is make clear by plotting \fdm-- and $\Mdyn/\mst$--\Re\ relations.

The  red   curves  in  Fig.   \ref{fig:fig2}  also   show  how
median \fdm\ trends  change when  one excludes objects  with
$\Mdyn  < \mst$, i.e. $\fdm  < 0$. Except  for the trend  with $n$
where a  small rigid offset is seen  ($\sim 0.05$ for the \SIS),
all correlations are only affected at one  end, i.e. at low mass,
velocity dispersion, and \Re, and at high \rhostar, where our
sample turns out to be incomplete (see above).   Hence, the  issue
of  negative \fdm\  values only  produce a slight  effect on  the
slopes  reported in  Tab.~\ref{tab:tab1}.

  Notice  that negative  \fdm\  values can  be  caused by  measurement
  errors  on galaxy  parameters  (i.e.  $\sigma$,  \Re,  and \mst),  a
  failure of the mass model to estimate \Mdyn, and/or some systematics
  in the stellar mass estimates, such as the assumption of a given IMF
  (see Sec.~\ref{sec:noDM}).  In order  to assess if these effects can
  fully account for the negative  \fdm\ values  in our sample, for each \mst,
  we assign  a mock \Mdyn\ value,  according to the \Mdyn--\mst\  relation of
  Fig.~\ref{fig:Mdyn_Mstar}  (for   a  \SIS\  model).    Shifting  the
  \mst\ and mock \Mdyn\  values according to the estimated statistical
  uncertainties  on these quantities  ($\sim 0.1$~dex  and $0.18$~dex,
  respectively),  we   find  $\sim  2  \%$  of   data-points  to  have
  \Mdyn$<$\mst. This fraction rises to $\sim 12 \%$ (i.e., what we
  actually  measure  for  the  \SIS\  model), if  stellar  masses  are
  overestimated  by a  factor  of $1.6$,  which is  within  the factor of $2$
  systematic   uncertainty   on  \mst\   due   to  different   effects
  (i.e. mainly  the extinction law and  IMF; see Sec.~\ref{sec:Smasses}
  and Paper  V).

  An alternative explanation for  the unphysical \fdm\ values is
  the choice of the galaxy  mass profile, which could be inappropriate
  for (some) low mass  galaxies.  For instance, dynamical masses would
  become systematically  larger (than the  \SIS\ ones) if  a power-law
  density profile, $\rho(r) \propto r^{-\alpha}$ with $\alpha < 2$, is
  adopted  (with $M(r)  \propto  r^{3-\alpha}$, i.e.   a mass  profile
  shallower than the \SIS). Different observations have argued that the (dynamical) mass
profile of a galaxy might change as a function of its stellar
mass. In particular, as already pointed out in Sec.
\ref{sec:Dmasses}, massive ETGs are quite well reproduced by
nearly-isothermal profiles (e.g. \citealt{Kochanek91, SLACS3};
\citealt{SLACSIV}), while in some intermediate-mass ETGs the mass
is found to follow the light
(\citealt{R+03,Nap+05,Douglas+07,Deason12}). Similarly,
non-homology of the mass profiles has also been suggested by T+09.
Qualitatively, this would make the trends of \fdm\ in terms of
\Re, $\sigma$, \mst\ and \Mdyn\ steeper with respect to the case
of a pure \SIS\ model, and shallower (or inverted) for the cases
of \fdm\ versus stellar density and S\'ersic index.

\begin{figure*}
\psfig{file=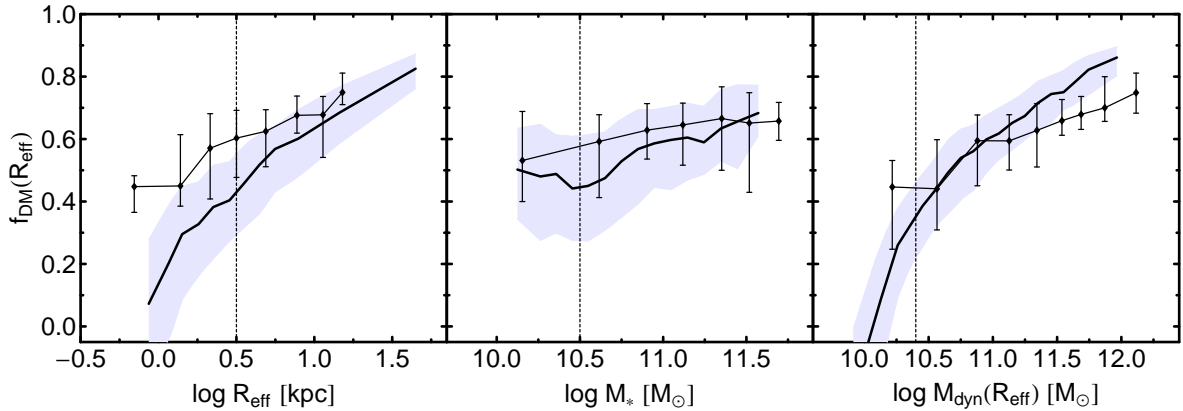, width=0.95\textwidth}\caption{\fdm\ within
$g$-band \Re\ in terms of \Re, \mst\ and \Mdyn\ for a \SIS. Solid
curves and shaded regions are for our sample, as in Fig.
\ref{fig:Mdyn_Mstar}, while points with error bars are from T+09
and NRT10. The vertical dotted lines mark the completeness limits
as in Fig. \ref{fig:fig2}.} \label{fig:fig_fDM_comparison}
\end{figure*}
Whilst most of these results are qualitatively  consistent with
those found in the (recent) literature,  one  should notice that,
in contrast to previous studies, our DM  fractions are  computed
within  the $K$-band (rather  than optical) effective radius.
Hence, our \fdm\ estimates are less affected by the existence of
metallicity and age gradients in ETGs, dust  extinction and (low
fractions of) young stars.
In  order to permit a more direct
comparison with results from our previous work (T+09),  we
recomputed  \SIS\ \fdm\ estimates  using $g$-band (rather than
$K$-band)  \Re\ values. Fig. \ref{fig:fig_fDM_comparison} compares
the resulting $g$-band median trends with the $B$-band trends
obtained from T+09. For the trend with stellar mass, we find good
agreement between the present and our previous work. However, the
plots of \fdm\ vs. \Re\ and \Mdyn\ reveal some  discrepancies. At
low \Re\ and \Mdyn, the present sample becomes incomplete and we
miss galaxies with high \fdm. This biases our trends towards lower
\fdm\ values with respect to T+09. At high \Re, our \Re\ values
are larger than those of T+09 (because of the different way
structural parameters are estimated; see
App.~\ref{sec:str_params}), explaining the shift of the \fdm\ vs.
\Re\ trend towards larger \Re\ with respect to T+09.
 On the other hand, the present \fdm\ vs. \Mdyn\ trend is shifted towards lower
\Mdyn\ values relative to T+09, implying that some further systematic
differences exist between the \Mdyn\ estimates of the two
datasets. Understanding the origin of such differences is far from
trivial as we do not have galaxies in common between the two
samples. Differences might exist because of different sample
selection criteria, and/or different methods used to estimate
velocity dispersions.

\begin{figure*}
\psfig{file=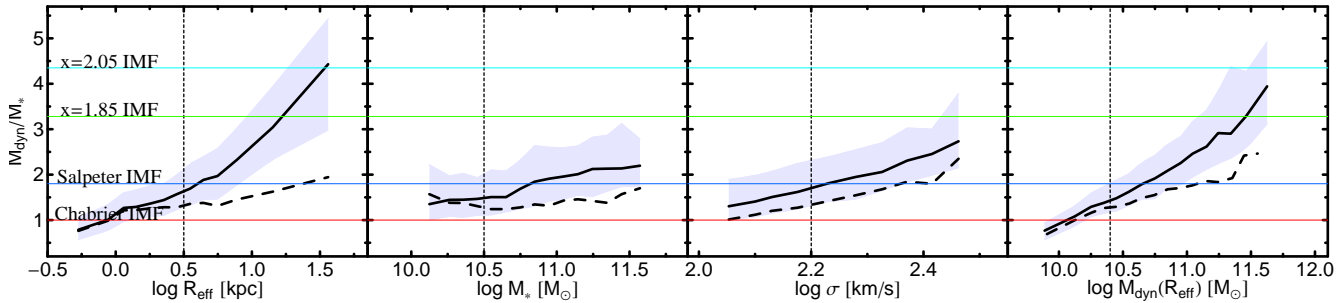,  width=1\textwidth}\caption{Trends  of
    \Mdyn/\mst\   as   a  function   of   $K$-band   \Re,  stellar   mass
    \mst\  (assuming   a  Chabrier  IMF),   velocity  dispersion,  and
    dynamical mass, \Mdyn. Black curves  and grey regions are the same
    as in Figs.   \ref{fig:Mdyn_Mstar} and \ref{fig:fig2}.  Horizontal
    lines  correspond  to  the  relative variation  of  stellar  mass,
    $M_{\star,\rm IMF}/M_\star$ --  with respect to a Chabrier IMF  -- when adopting
    different IMFs,  with slopes $1.35$  (i.e. a Salpeter  IMF; blue),
    $1.85$ (green), and $2.05$ (cyan). The red line corresponds to the
    case  of a Chabrier  IMF ($M_{\star,\rm IMF}=M_\star$).   The vertical
    dotted   lines   set   the   completeness  limit   as   in   Figs.
    \ref{fig:fig2}      and     \ref{fig:fig_fDM_comparison}.      For
    \fdm$=M_{\star,\rm IMF}/M_\star$,  one  can find  the  IMF slope  that
    would make \mst\  equal to \Mdyn, resulting in a galaxy with no
    DM.  } \label{fig:noDM}
\end{figure*}

\subsection{Constraints on IMF}\label{sec:noDM}

The present analysis is  based on the assumption that the stellar
  IMF is universal,  at least within the parameter  range covered by
  our sample.  In  principle, for each given galaxy,  one could adjust
  the IMF  so that  \mst=\Mdyn\ (or, equivalently,  \fdm$=0$).  However,
  since  \fdm\  correlates with  various galaxy  parameters, any
  adjustment of the IMF would imply  that the IMF itself changes
  systematically  with these parameters, in  the same  way  as \fdm\ (\citealt{Cappellari+12}).
  Motivated  by  the recent  claim  that  the  IMF depends  on  galaxy
  velocity   dispersion~\citep[hereafter  vDC11]{vDC11},   with  high-
  (relative to low-) $\sigma$ ellipticals having a bottom-heavier IMF,
  we explore to what extent one can change the IMF in order to fulfill
  the  condition  \fdm$=0$.

  We  estimate  the  expected variation  of
  stellar mass,  $M_{\star,\rm IMF}/M_\star$, relative to  a Chabrier IMF,
  when  different IMFs are  adopted.  Here,  $M_\star$ is  the stellar
  mass  estimated with a  Chabrier IMF,  while $M_{\star,\rm IMF}$  is the
  stellar mass we  would estimate with a different  IMF.  In practice,
  we  consider   three  power-law  IMFs,  with   slopes  $1.35$  (i.e.
  Salpeter),  $1.85$, and $2.05$  (i.e.  a  very bottom-heavy
  IMF,  as  suggested by~vDC11  for  high-$\sigma$ ellipticals).   The
  $M_{\star,\rm IMF}/M_\star$  is estimated  as the  ratio of  the stellar
  $M/L$  between two  SSPs having  a  power-law and  a Chabrier  IMFs,
  respectively.
  To compute the stellar $M/L$\footnote{We compute the $M/L$ values
    in  the $K$-band, as this is less  sensitive to  the presence  of young
    stars and metallicity of a stellar population.}, we adopt the BC03
  synthesis  code, for  old ($10$~Gyr)  SSPs, with  solar metallicity.

  Fig.~\ref{fig:noDM}  plots the  \Mdyn/\mst\ trends  as a  function of
  \Re, \mst, \Mdyn, and $\sigma$ (black curves), with horizontal lines
  marking  $M_{\star,\rm IMF}/M_\star$   for  the  different   IMFs.   The
  intersections of  the horizontal lines with the  black curves define
  the values of \Re, \mst, \Mdyn,  and $\sigma$, for which a given IMF
  slope  would imply  \fdm$=0$.  The  Figure  shows that  in order  to
  account for the apparent trend of \fdm\ with \Re\  (and \Mdyn), galaxies with
  the largest radii  (and dynamical  masses) should have  an IMF  slope as
  steep as $2.05$, while at the lowest \Re\ (and \Mdyn) a Chabrier (or even
  a  bottom-lighter) IMF  would be  required.  Interestingly,  at high
  velocity dispersion ($\log  \sigma \sim 2.45$), we see  that the IMF
  slope cannot be larger than  $\sim 1.8$, which  seems to contrast with
  the finding of~vDC11, but is in good agreement with \cite{Cappellari+12} (see their no-DM case).

The exercise  described so far shows that the
  estimated trends of  \fdm\ might be significantly biased  if the IMF
  varies substantially  as a function of  different galaxy parameters.
  However,  it remains  unclear  if and  (more  importantly) why  such
  correlated variation  of IMF slope with  different parameters should
  be present.   Another key to  the \fdm\ trends might  be alternative
  gravity       (\citealt{Cardone+11_MOND};       \citealt{Lubini+11};
  \citealt{Richtler11,Nap+12}), which, anyway,  is beyond the scope of
  the present work.

\subsection{Dark matter densities}\label{sec:dmdens}

Following  T+09 and  NRT10, we  also  analyze the  average central  DM
density of  ETGs, \rhoDM.   The \rhoDM\ values  may be  considered the
fossil  record of  the ambient  density at  the time  of  initial halo
collapse, modulo possible  mass-redistributing interactions between DM
and  baryons.  While  \fdm\ is  defined relative  to the  baryon mass,
\rhoDM\ quantifies the properties  of the DM component alone, allowing
for   more   direct    comparisons   to   cosmological   models   (see
Sec.~\ref{sec:DM_models}).

\begin{figure*}
\psfig{file=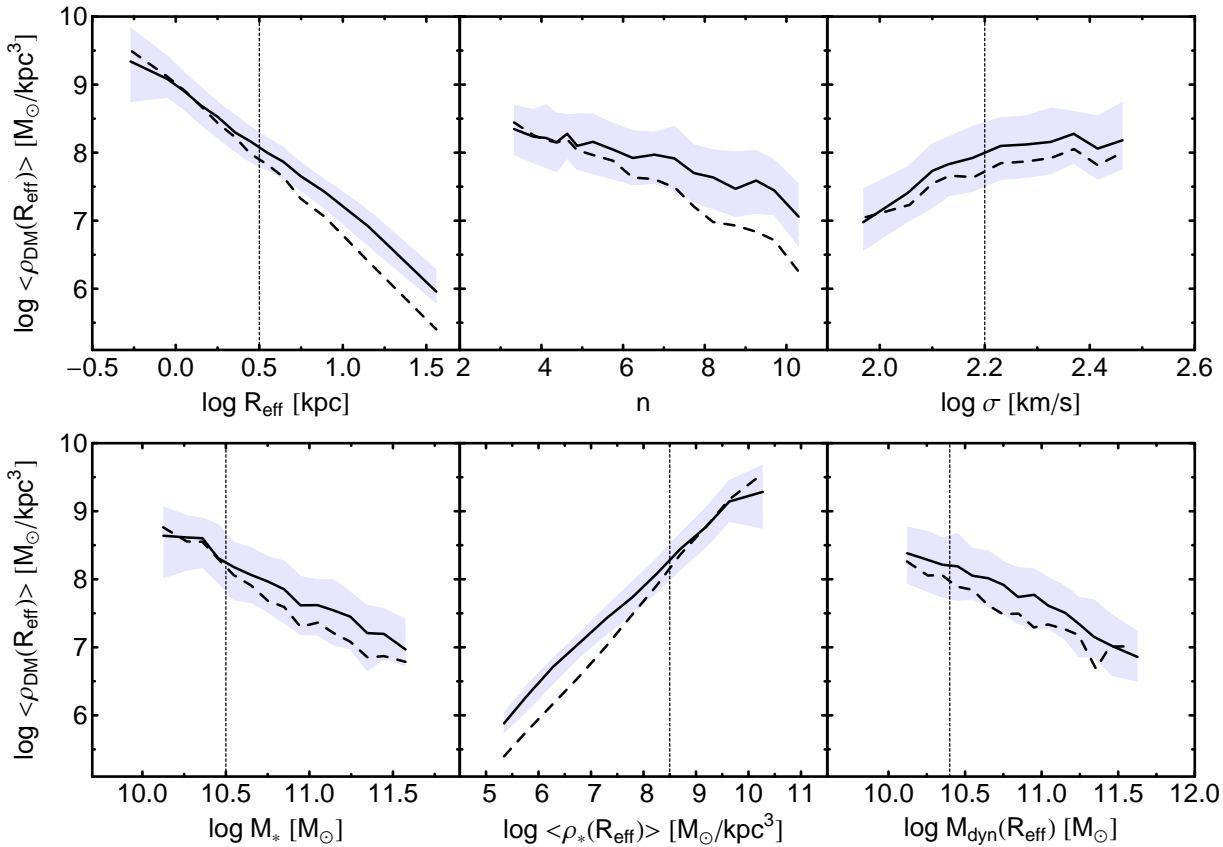, width=1\textwidth}\caption{Average central
DM density as a function of $K$-band \Re, $n$, velocity dispersion
$\sigma$, stellar mass \mst, central average stellar density
\rhostar\ and dynamical mass, \Mdyn. The symbols are as in Figs.
\ref{fig:Mdyn_Mstar} and \ref{fig:fig2}, except as for the red
curves in that plot, here only the galaxies with $M_{\rm dyn}
> \mst$ are used. The vertical dotted lines set the completeness
limit as in Figs. \ref{fig:fig2} and \ref{fig:fig_fDM_comparison}.
} \label{fig:fig3}
\end{figure*}

Fig.~\ref{fig:fig3} is the same as  Fig.~\ref{fig:fig2}, but
plotting \rhoDM, rather than \fdm, as a function of  structural
parameters and galaxy mass proxies.  The correlations are again
fitted with linear relations, taking into account the completeness
limit of each diagram, with the slopes reported in Table
\ref{tab:tab1}.
We find that \rhoDM\ decreases with  \Re, $n$, \mst, and \Mdyn.
The best-fitting power laws are consistent with results from the
literature (\citealt{Thomas+09}; T+09; NRT10; \citealt{SLACSX};
T+10). On the other hand, \rhoDM\ increases with $\sigma$ and
\rhostar. All trends are qualitatively independent of the  adopted
mass model, although in general the \cMtoL\ profile  produces
steeper correlations than the \SIS. As shown in
Fig.~\ref{fig:fig3}, at low \Re,  $n$, $\sigma$, \Mdyn, and high
\rhostar, \rhoDM\ is model independent, while the discrepancy
between \SIS\ and \cMtoL\ models is larger at high \Re,  $n$,
$\sigma$, \Mdyn, and low \rhostar, making the  \cMtoL\
correlations steeper. This is due to the increasingly
important role of mass profile extrapolation in these regimes.

Under the  assumption that DM density profiles  are universal, at
  least  for all galaxies  in our  samples, we  confirm the  result of
  NRT10, that DM  profiles in ETGs are very  cuspy, with $\rho_{DM}(r)
  \sim  r^{-2}$.   This conclusion  is  reached  as  follows.  For  a
  power-law    DM   density   distribution,    $\rho_{DM}(r)   \propto
  r^{-\alpha}$, we have  $M_{DM} = r^{3-\alpha}$ (for $\alpha  < 3$ ).
  It  follows that $\rhoDM(\Re)  \propto  \Re^{-3}   M_{DM}(\Re)  \propto
  \Re^{-\alpha}$.  Since  the slope of the \rhoDM\  -- \Re\ correlation
  is $\sim 2$ for both \SIS\  and \cMtoL, we conclude that $\alpha \sim 2$,
  independent of the adopted mass model. This conclusion relies on the
  assumption  that trends  in the  average values  of  \rhoDM\ reflect
  trends   in  the   mass  profiles   of  individual   galaxies  (cf.
  \citealt{Walker09}; NRT10).  We  will return to further implications
  of this assumption in Section~\ref{sec:DM_models}.

\section{``Dark matter plane'' of ETGs}
\label{sec:DM_plane}
\begin{table}
\centering \caption{Comparison of the $\fdm-X$ and $\fdm-\Ymed$
relations (black and red curves in Fig.~\ref{fig:fdm_corr}; see
the text). We flag with `Yes' (`No') those cases where
$\fdm-\Ymed$ overlaps (does not overlap) with
$\fdm-X$.}\label{tab:tab2}
\begin{tabular}{lccccc}
\hline
 & \multicolumn{4}{c}{$Y'$} \\
 & $\Re$  & $\mst$ & $n$ & $\sigma$ & $\rhostar$  \\
 \hline
 $X=\log \Re$      & $-$ & No & No & No & Yes    \\
 $X=\log \mst$     & No & $-$ & No & Yes & Yes    \\
 $X=n$    & Yes & No & $-$ & No & Yes    \\
 $X=\log \sigma$   & No & No & No & $-$ & No    \\
 $X=\log \rhostar$ & Yes & No & No & No & $-$    \\
  \hline
\end{tabular}
\end{table}
\begin{figure*}
\psfig{file=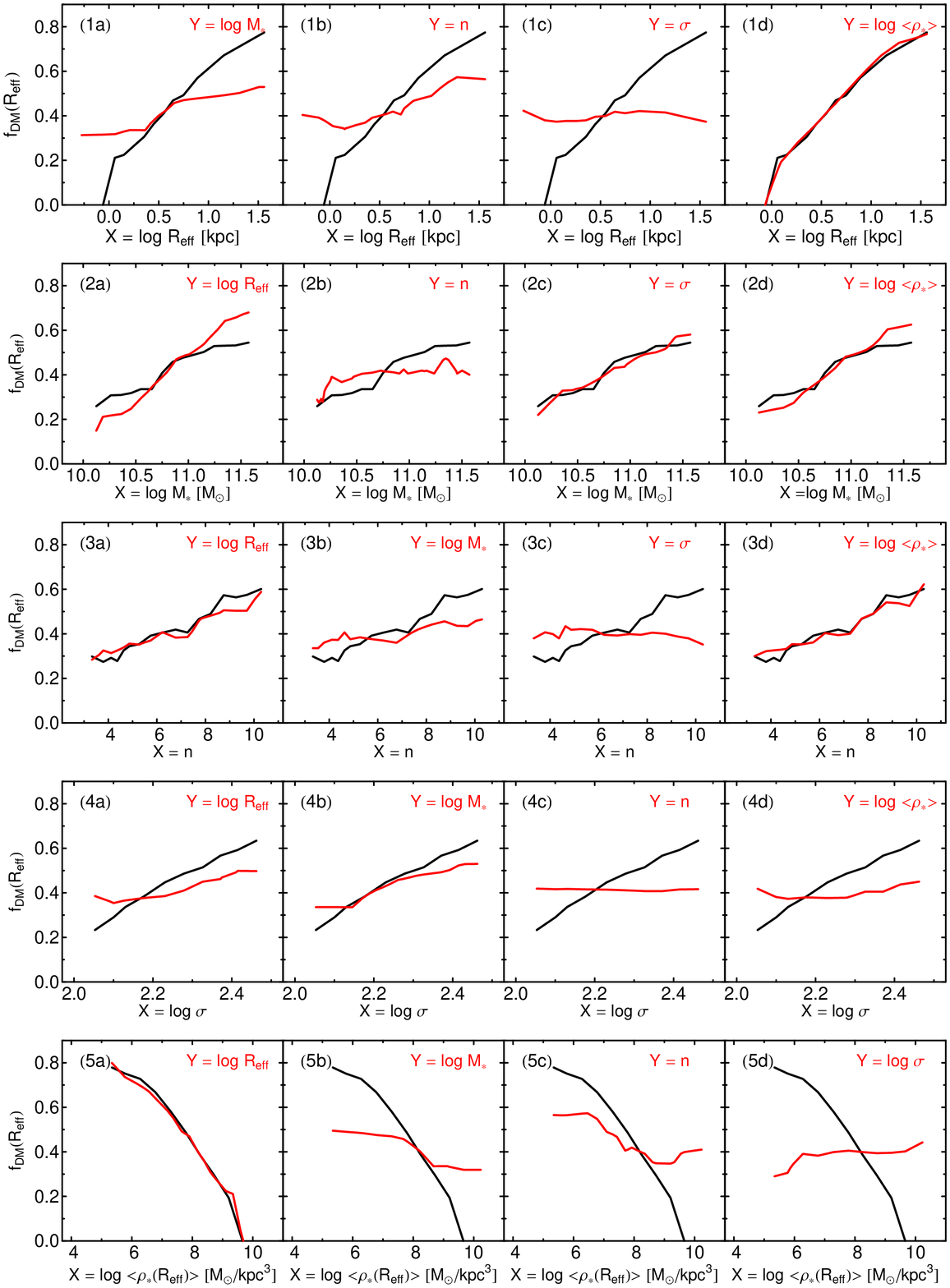, width=15cm} \caption{ Tests  performed to
establish the  observed ETGs' parameters
  driving the correlations with central DM fraction.  Each panel shows
  the  observed correlation between  \fdm\ and  a given  parameter $X$
  (black solid curve),  as well as the correlation  expected from that
  of \fdm\ and another galaxy parameter $Y$ (i.e.  $Y \ne X$), and that
  between  $Y$  and $X$  (red  solid curve).   Each  row  in the  plot
  corresponds to a given $X$,  out of \Re, \mst, $n$, $\sigma$,
  and \rhostar\ (from top to bottom). For a given row, the panels
  show  the same \fdm--$X$  correlation (black)  and that  expected by
  varying $Y$ (from left to right).  For each panel, the corresponding
  $Y$ is reported in the upper--right corner.  }
\label{fig:fdm_corr}
\end{figure*}
Some of the correlations discussed so far, between DM fraction
(density) and different galaxy parameters, could in principle be
secondary in nature, arising simply from mutual correlations among
the galaxy parameters themselves.   For  instance,  given the
\fdm\--\Re\  relation,  a correlation  between \fdm\ and stellar
mass  is also expected to exist,  just  because of  the
size--stellar mass relation (e.g.~\citealt{Shen2003}).   Therefore
we seek to determine the fundamental parameters that relate to the
DM matter  content  of ETGs.
We  follow the  same procedure adopted in  \cite{LaBarbera+10d}
(Paper  IV)  to  establish the main drivers  of correlations among
internal   colour gradients and other observed parameters of ETGs.
As illustrated in Fig.~\ref{fig:fdm_corr}, we consider a given
galaxy parameter, $X$, out of the quantities $\{$\Re, $n$,
$\sigma$, \mst, and $\rhostar$$\}$ (from top to bottom in the
Figure),   and the corresponding correlation with \fdm, $\fdm(X)$
(black curves in each panel).  For a given bin of  $X$, we compute
the median value ($\Ymed$) of another galaxy parameter, $Y$, with
$Y \ne X$, and then, given the \fdm$(Y)$ correlation, we compute
$\fdm(\Ymed)$ (red  curves in the Figure). The $\fdm(\Ymed)$ is
the correlation between \fdm\ and $X$ expected from the
correlation between \fdm\ and $Y$, and the correlation between $X$
and  $Y$ themselves.
Each row in Fig.~\ref{fig:fdm_corr}, from top to bottom,
corresponds to a given $X$,  while the panels from left to right
show  the $\fdm(\Ymed)$ (see red curves) by varying $Y$. If a
given $\fdm(\Ymed)$  differs  from  $\fdm(X)$ (i.e. the black and
red curves in the corresponding panel do not overlap),   we  can
conclude  that  the correlation  between \fdm\  and $X$  is more
than just a reflection of one between \fdm\ and $Y$. We evaluate
by visual inspection whether or not two curves differ, as
summarized in Table \ref{tab:tab2}, for each pair of $X$ and $Y$.
We find that the
correlation between \fdm\ and $X=\log \mst$ (panels 2a--2d) is
equivalent to that expected from either \fdm\ and $\sigma$ (panel
2c), or \fdm\ and \rhostar\ (panel 2d), i.e. stellar mass is
likely not a genuine driver of the DM fraction in ETGs. In the
same way, the correlation with S\'ersic $n$  is equivalent to that
expected between \fdm\ and \Re\ (or $\log \rhostar$) (panels 3a
and 3d). On  the contrary, the correlation between \fdm\ and
$\sigma$ is a genuine one,  as all the  red curves  in panels
4a--d  of Fig.~\ref{fig:fdm_corr} differ, being flatter than  the
observed \fdm$(\log \sigma)$.  Finally, we see that the \fdm$(\log
\Re)$ and \fdm$(\log \rhostar)$ correlations are completely equivalent to each
other~\footnote{In fact, the observed \fdm$(\log \Re)$ relation is fully
consistent with that expected from \fdm$(\log \rhostar)$ (panel
1d), and vice-versa (panel 5a).}, and are not caused by any
correlation of \fdm\ with some other parameter. Since $\rho_\star$
is computed from $\Re$, we conclude that the main parameters
driving DM fractions  are galaxy size and central velocity
dispersion.  The  same result holds when repeating the above
analysis using DM density, rather than \fdm.

\begin{figure}
\psfig{file=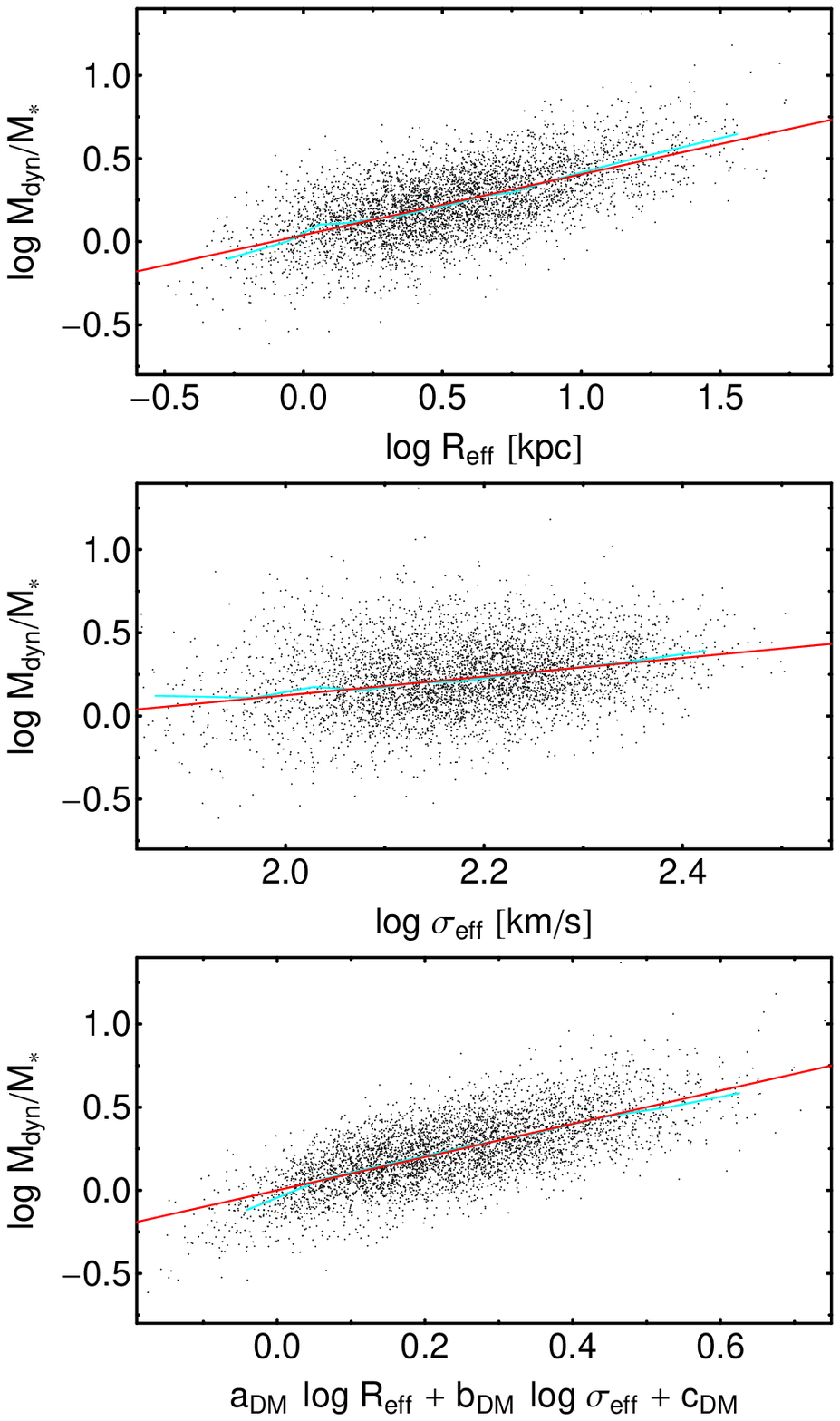, width=0.44\textwidth} \caption{DM
fundamental plane. {\it Top Panel.} $\log \Mdyn /\mst$ as a
function of \Re. {\it Middle Panel.} $\log  \Mdyn /\mst$ as a
function of $\sigma_{e}$. {\it Bottom Panel.} $\log \Mdyn/\mst$ as
a function of the combination $a_{\rm DM} \log \Re + b_{\rm DM}
\log \sigma_{e} + c_{\rm DM}$. The cyan lines are the medians,
while the red ones are the best fits.} \label{fig:DM_plane}
\end{figure}

Since there are two observed parameters driving the DM content of
ETGs (\Re\ and $\sigma$),  we analyze here the correlation  of DM
with both quantities,  using  \SIS--based  DM  estimates.  Here,
we adopt  the logarithmic ratio of  dynamical to stellar mass,
$\log\,(\Mdyn/\mst)$. This avoids the issue of negative \fdm\
values and can provide a more direct connection to the  FP
relation (see  \S~\ref{sec:FP_tilt}).
We consider a DM plane relation:
\begin{equation}\label{eq:DM_plane}
\log \frac{M_{\rm dyn}}{\mst} = a_{\rm DM} \log \Re +b_{\rm DM}
\log \sigma_{e} + c_{\rm DM},
\end{equation}
where $a_{\rm DM}$  and $b_{\rm DM}$ are the  slopes, and $c_{\rm DM}$
is the offset. We perform a linear best-fit of $\log\,(\Mdyn/\mst)$ vs. $\log \Re$
and  $\sigma(\Re)$, minimizing the sum  of  absolute residuals along the
$M_{dyn}/\mst$ axis. This gives $a_{\rm DM} =0.380 \pm
0.008$, $b_{\rm  DM} = 0.647 \pm  0.03$, $c_{\rm DM}  =-1.38 \pm 0.06$
(where errors are $1  \sigma$ statistical uncertainties).
Both $a_{\rm DM}$ and $b_{\rm DM}$ are significantly different from
zero, implying that DM content is indeed a function of both \Re\
and $\sigma$, consistent with what we found above (i.e. that the
correlation of \fdm\ and \Re\ is not equivalent to that with
$\sigma$, and vice-versa).
Fig.~\ref{fig:DM_plane} compares  the (logarithmic) correlations
of   $M_{dyn}/\mst$  vs. $\Re$  (top), $M_{dyn}/\mst$   vs.
$\sigma(\Re)$   (middle),  and those between $M_{dyn}/\mst$ and
both  $\Re$ and  $\sigma(\Re)$ (i.e. the DM plane; see bottom
panel). See the slopes reported in Table \ref{tab:tab1}. For all
panels, the red lines are the best-fitting relations obtained by
the same kind of fitting procedure, i.e. minimizing the sum  of
absolute residuals along the $M_{dyn}/\mst$ axis. A robust
estimate of the rms along the $M_{dyn}/\mst$ axis is also reported
in the Figure for each correlation. Notice that the scatter of the
DM  plane ($\rm rms = 0.149 \pm 0.002$) is smaller (albeit  by a
few percent) than that of the $M_{dyn}/\mst-\sigma(\Re)$   ($\rm
rms = 0.200 \pm 0.002$) and $M_{dyn}/\mst-\Re$ ($\rm rms =
0.165\pm 0.002$) correlations.

Projecting the $M_{dyn}/\mst$ rms along the $\log \Re$ axis, we find a
scatter  of $\sim  0.4$~dex,  i.e. much  larger  than that  of the  FP
relation ($\sim 0.1$~dex; see Paper II). This large dispersion implies
that  the coefficients  of the  DM plane  depend significantly  on the
fitting procedure  one adopts to derive them.   For instance, minizing
the rms  of absolute residuals along  $\log \Re$, we  find $a_{\rm DM}
\sim  0.84$  and  $b_{\rm   DM}\sim  0.78$,  which  are  significantly
different from  those obtained  from the $M_{dyn}/\mst$  best-fit (see
above). The scatter of the DM plane decreases to $\sim 0.25$~dex along
$\log \Re$, still significantly larger with respect to the FP.

\section{Dark matter fractions and Fundamental Plane}\label{sec:FP_tilt}
 The correlations of DM content with galaxy parameters might help
to shed light on the  origin  of  scaling  relations  of ETGs,
i.e. the tilt  of the FP relation. The FP can be written as
\begin{equation}
\log \Re =  a_{FP} \log \sigma + b_{FP}  \langle \mu_\star
\rangle_e + \, constant,
\end{equation}
where $a_{FP}$ and $b_{FP}$ are the slopes, and $\langle \mu_\star
\rangle_e$ is the mean surface brightness within \Re. Under the
assumption of homology, the tilt can be parameterized as a
variation of $\Mdyn/L$ with luminosity or mass, i.e. $\Mdyn/L
\propto L^{\alpha}$ (\citealt{Dressler87}), or $\Mdyn/L \propto
{\tt \Mdyn}^{\gamma}$ (with $\gamma=\alpha/(1+\alpha)$). The \ML\
can be rewritten as
\begin{equation}
\frac{\Mdyn}{L} \propto \frac{\Mdyn}{\mst} \frac{\mst}{L},
\end{equation}
where $\mst/L$  is the stellar mass-to-light ratio.   In the $K$-band,
stellar population effects make  a negligible contribution to the tilt
(\citealt{LaBarbera+10b}),   implying  that   non-homology   and/or  a
variation   of   DM   content   (i.e.   $\Mdyn/\mst$   changing   with
luminosity/mass)  should explain  the tilt.  Fig.~\ref{fig:fig2} shows
that the central DM content of ETGs does actually change significantly
with mass  and luminosity. In particular, we  find $\Mdyn/\mst \propto
\Mdyn^{0.36   \pm    0.03}$,   or   $\gamma=0.36    \pm   0.03$   (see
Table~\ref{tab:tab1}). This  value is somewhat larger  than that found
by   some   previous  studies,   i.e.   $\gamma  \sim   0.15$--$0.20$,
(\citealt{HB09},  \citealt{Gallazzi+06}  and  T+09), possibly  due  to
differences  in  the  way  structural parameters  are  estimated  from
different  sources  (see  App.~\ref{sec:str_params}).   The  value  of
$\gamma$,  as  obtained  by  fitting  the  $\Mdyn/\mst$  vs.   $\Mdyn$
relation, is  close to that  inferred from FP coefficients,  using the
virial theorem  under the assumption  of homology ($\gamma  \sim 0.2$;
see, e.g., T+09; \citealt{LaBarbera+10b}).  This means that DM content
variation might  be able by itself  to explain the  entire fraction of
the  tilt which  is not  due to  stellar population  effects  (i.e.  a
variation of age and/or metallicity  with mass).  We also notice that,
in the context of the \cMtoL\  model, the variation of DM content with
mass is  flatter than with  the \SIS\ model,  which would lead  one to
conclude that both non-homology and  a change of DM content contribute
to the tilt  (see T+09 for details).  Last, but  not least, one should
notice  that connecting  the  slope of  the  $\Mdyn/\mst$ vs.  $\Mdyn$
relation to  the tilt  of the  FP is not  trivial. The  dynamical mass
entering the  virial theorem is actually  the total mass  of a galaxy,
i.e. all the  gravitationally bound matter of the  system.  Hence, the
difference  of FP  coefficients  from the  virial theorem  expectation
might be more related to the total, rather than central ($<1$~\Re), DM
content  of  ETGs.    Unfortunately,  estimating  total,  rather  than
central, DM  fractions requires a  large extrapolation of  the central
velocity dispersion, much more dependent on the mass model one assumes
to infer our default $\Mdyn$ values.  With this caveat in mind, we try
here  to establish  a  connection between  the  DM plane  and the  FP,
combining Eq.~(\ref{eq:DM_plane}) with the following expression of the
virial  theorem, obtained  under the  assumption of  homology (whereby
kinetic energy  is $\propto M_{dyn} \sigma^2$ and  potential energy is
$\propto M_{dyn}^2/\Re$):
\begin{equation}
\sigma^2 \propto G  \frac {\Mdyn}{\mst} \frac{\mst}{\Re},
\end{equation}
where $G$ is the  gravitational  constant.   We  obtain an
equation formally identical to the FP:
\begin{equation}
\log \Re =  a \log \sigma + b \langle \Sigma_\star \rangle_e + \,
constant,
\end{equation}
with $a=(2-b_{DM})/(1+a_{DM})$, $b=0.4/(1+a_{DM})$, and $\langle
\Sigma_\star \rangle_e$ is the mean  stellar mass density  within
\Re.

\begin{figure*}
\psfig{file=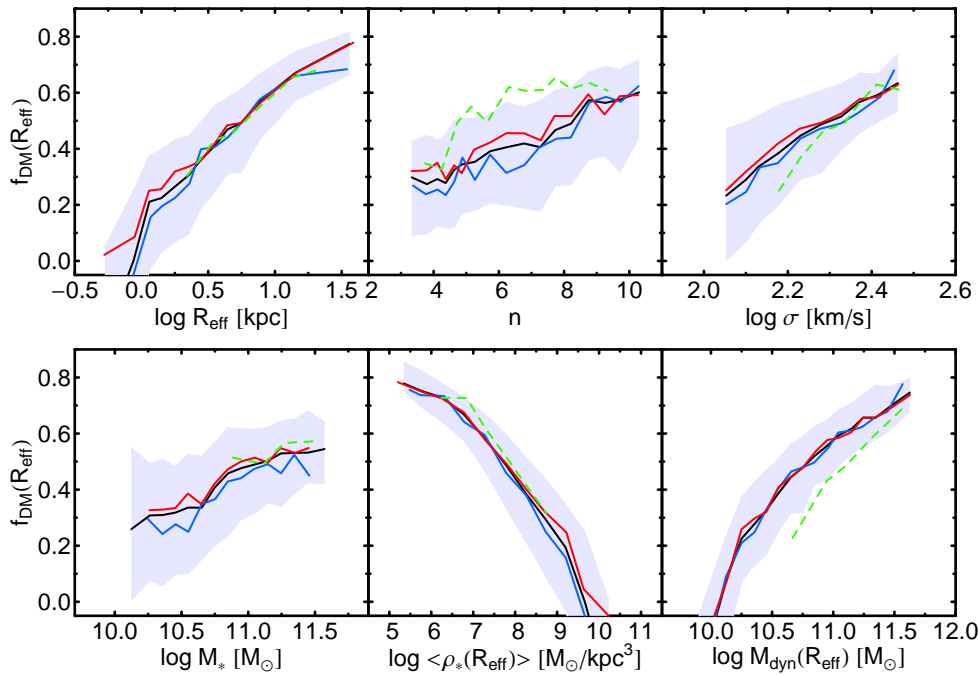, width=0.80\textwidth}\caption{Environmental
dependence of the correlations between DM   fraction   and
structural parameters/mass proxies. Only the \SIS\ model is shown.
Solid black lines and  shaded regions  are the same as in
Fig.~\ref{fig:Mdyn_Mstar}, and refer to the entire sample. Blue,
red, and green lines  show the  median trends  for field,
satellite, and central galaxies, respectively.}
\label{fig:fig_env}
\end{figure*}

Using     the    values    of     $a_{DM}$    and     $b_{DM}$    from
Sec.~\ref{sec:DM_plane},  we obtain  $a=1.00 \pm  0.02$,  $b=0.290 \pm
0.002$.   These  coefficients can  be  compared  to  those derived  by
~\citet{HB09b} for  the stellar mass  FP, i.e. $a_\star \sim  1.4$ and
$b_\star \sim  0.32$ (when  using a direct  fitting method,  see their
tab.~3),  and  those we  derived  for the  $K$-band  FP  in Paper  II,
i.e.  $a_K  \sim  1.55$  and  $b_K  \sim 0.32$.   While  $b$  is  very
consistent  with  $b_\star$  and  $b_K$,  $a$  is  smaller  than  both
$a_\star$ and  $a_K$.  This discrepancy might be  explained by several
effects,  such  as  selection  effects (see  e.g.~Paper  II),  fitting
procedure  (see   comments  about   DM  plane  fitting   procedure  in
Sec.~\ref{sec:DM_plane}),  and the  method  (i.e. the  model) used  to
estimate  DM fractions.  In  general, the  fact that  $a$ and  $b$, as
derived   above,  are  significantly   smaller  than   the  homologous
expectations  of the  virial  theorem (i.e.  $a=2$  and $b=0.4$),  and
similar (or even smaller) with respect to the observed coefficients of
the stellar mass (or $K$-band) FP,  may lead one to conclude that most
of the tilt  is due, indeed, to a variation of  DM content with galaxy
mass (or $\sigma$ and \Re,  as shown here).  However, we still remark,
as noticed  above, that  the DM plane  is a correlation  among central
quantities of ETGs (with dynamical mass being estimated within 1~\Re),
while the  virial theorem  is a global  relation (between  kinetic and
potential  energy  within  an  infinite  aperture).  Thus,  the  above
conclusion about  the origin  of the tilt  remains uncertain.   On the
other hand,  the existence of a  central--DM plane is  a robust result
(see previous section),  and may be explained by the  DM halos of ETGs
being only  approximately universal.  A larger \Re\  encloses a larger
portion  of  the  halo,  implying  \fdm\ to  increase  with  \Re\  (as
discussed by, e.g.,  NRT10).  At given \Re, a  larger $\sigma$ means a
deeper central potential  well, i.e.  more DM in  the centre, implying
that $\Mdyn/\mst$ increases with both \Re\ and $\sigma$.

\section{Central dark matter and galaxy environment}\label{sec:DM_env}

In  a hierarchical paradigm  of galaxy formation, the  (dark) matter content
of the outer, less bound  regions of a galaxy can be stripped off,
as the galaxy is accreted into  a bigger halo, becoming a satellite. If stripping is effective enough, one might
see some variation also in the central DM matter fraction of
galaxies,  as a function of the environment   wherein   they
reside.    Environment-driven interactions, such as (both
major  and minor) merging with other group members, may
further change the DM fractions of galaxies residing in  groups.
In  this  section  we investigate the  effect of  galaxy
environment    on    the     central    DM fractions    of ETGs.
Fig.~\ref{fig:fig_env}  plots   \fdm\  as  a   function of
both structural  parameters and mass proxies (as  in
Fig.~\ref{fig:fig2}), for  ETGs classified  as field  and group
galaxies, the  latter being splitted between satellites  and
centrals (see Sec.~\ref{sec:sample}). We show here only \SIS\ results,
as \cMtoL\ trends would not add any relevant information to the analysis.

\begin{figure*}
\psfig{file=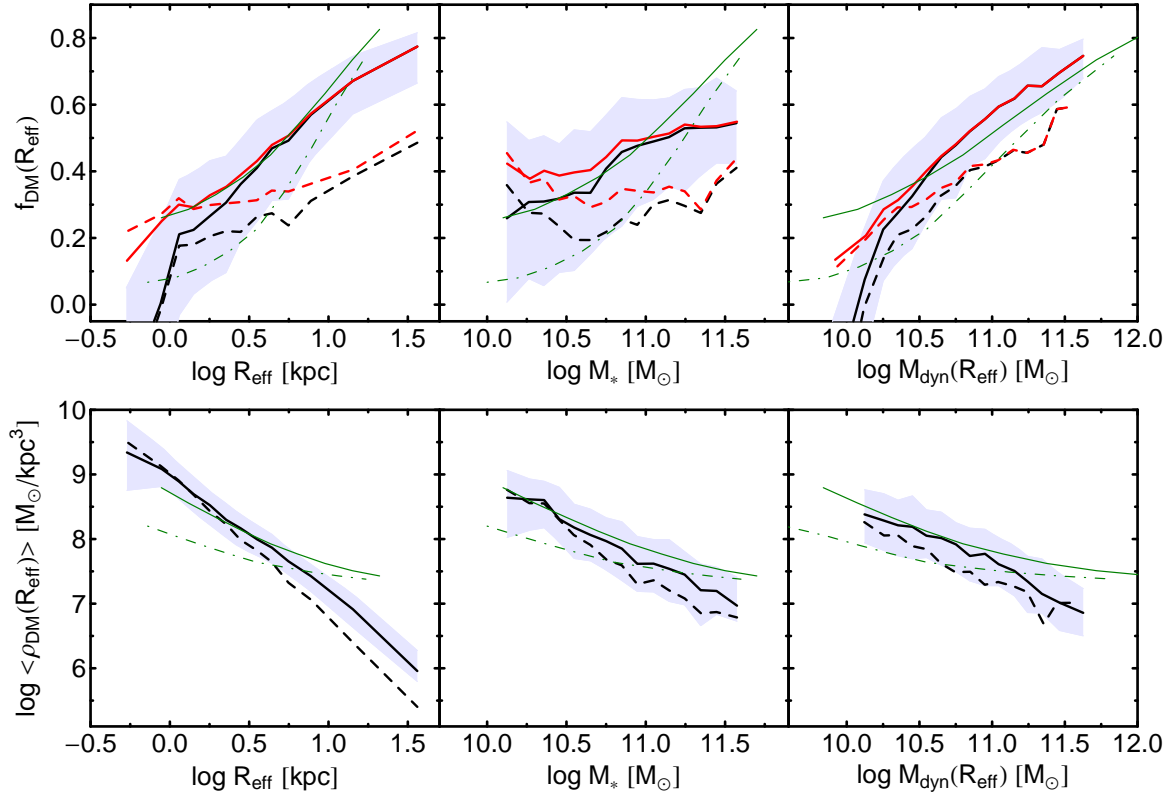,
  width=0.95\textwidth}\caption{Comparison of  the DM content  of ETGs
  to  the predictions  of galaxy-toy  models.  \fdm\  and  \rhoDM\ are
  plotted as  a function  of \Re,  \mst\ and \Mdyn\  in the  upper and
  lower panels,  respectively. Black and  red curves, as well  as grey
  shaded  regions are the  same as  in previous  plots. The  solid and
  dot-dashed green  lines are toy-model predictions  for a AC-NFW
  and NFW DM halo (see the text).}
\label{fig:fig_simulations}
\end{figure*}

We find that field  and satellite  galaxies  exhibit  fully
consistent \fdm\ trends while somewhat different  trends are
detected for centrals. In fact, when compared to  the average
trends of field  and satellite ETGs, centrals  have, at fixed  $n$
(\Mdyn), larger (lower) \SIS\  \fdm\ values; while  all remaining
correlations do not show any significant environmental dependence
(although at low $\sigma$ centrals also tend to have lower \fdm).
These differences  are just due to the fact that for centrals the
\mst\ (\Re) is  typically larger relative to field and satellite
galaxies, as seen by the \mst\ and \Re\  ranges for central
galaxies  (green curves)  in the \fdm--\mst\ and \fdm--\Re\
diagrams of Fig.~\ref{fig:fig_env}.  Since \fdm\  increases with
both \mst\ and \Re,  centrals tend to have, on  average, larger
\fdm, while at fixed \Mdyn,  a larger \mst\ implies a lower \fdm.
In order to account for a possibly different shape of the DM halo
of satellite (relative to field) ETGs, we have also computed their
\fdm\ values by  using  a truncated \SIS\ profile (see
Sec.~\ref{sec:Dmasses}). We find no significant difference with
respect to the results of a pure \SIS\ model.
We also test for environmental dependence in the DM plane of ETGs.
Minimizing the rms of absolute residuals along $\Mdyn/\mst$,
as done in Sec.~\ref{sec:DM_plane} for the entire sample, we obtain
$a_{\rm DM}=0.39 \pm 0.02$ ($b_{\rm DM}=0.58 \pm 0.05$) and $a_{\rm DM}=0.35 \pm 0.01$
($b_{\rm DM}=0.60 \pm 0.04$) for field and group ETGs, respectively, i.e. we still do
not detect any significant environmental dependence.
In summary, we do not  find any significant environmental
variation of the   correlations   between central   DM  content
and   structural parameters and mass of ETGs.  This result  can be
discussed in light of what we found in  Paper III,  i.e. that  the
tilt  of the  FP is  larger  for group, relative to field, ETGs.
Since we found this effect to be independent of the waveband where
galaxy structural parameters are measured (from $g$  through
$K$),  we  concluded   that  it  is  explained  by  some
``wavelength-independent''  effect, like a  different variation
of DM content with  mass for  group (relative to  field) ETGs.
This latter conclusion   seems  to  contrast   with  what   we
have   found  here (Fig.~\ref{fig:fig_env}).      However,    as
pointed     out    in Sec.~\ref{sec:DM_plane}, in  the present
study we analyze  central DM fractions, while  the tilt of the FP
may be  more related to the  behavior of global  quantities.
Alternatively,  the environmental variation of  the tilt  should
be explained  by a change  of dynamical structure   (i.e.
non-homology)  rather   than   DM  content   with environment.

\section{Cosmological models and simulations}
\label{sec:DM_models}

In order  to interpret  the observational results,  we follow  the
same approach as in  T+09, NRT10 and T+10, constructing a set of
toy-galaxy mass models,  whose DM density profiles are based on
$\Lambda$CDM cosmological simulations, i.e.   a \citet[hereafter
NFW]{NFW} profile.  A different set of models is also constructed
by applying a suitable recipe for  adiabatic contraction (AC) from
baryon settling to the  NFW profiles~\citep{Gnedin+04}.  In both
cases (i.e.  NFW and  AC-NFW models), a de-projected S\'ersic law
is used to describe the density profile of the stellar component.

The NFW halos are assumed to follow an average mass-concentration
relation, as in \cite{Nap+05},  while the S\'ersic  profiles are
assumed to follow the \Re-\mst\ and $n-\mst$ relations from our
data\footnote{ The $\Re-\mst$ relation is well reproduced by the
best fitted relation $\log \Re = -8.7 + 0.86 \, \log \mst/\Msun$,
while is $\log n = -5.3 + 0.58 \, \log \mst/\Msun$ and $\log n =
0.27 + 0.05 \, \log \mst/\Msun$ for galaxies with $\log \mst/\Msun
\lsim 10.5$ and $\gsim 10.5$.}. The virial DM mass is
parameterized in terms of a star formation efficiency
$\epsilon_{\rm SF}=M_\star/(\Omega_{\rm bar} M_{\rm tot})$, where
$\Omega_{\rm bar}=0.17$ (\citealt{WMAP2}) is the baryon density
parameter and $M_{\rm tot}$ is the total mass.  We assume that
$\epsilon_{\rm SF}$ varies with \mst, following the recipes
of~\citet{CW09}, where \eSF\ decreases from a maximum of $\sim
0.19$  at $\log$\,\mst$\sim 10.4$  down to $\sim 0.004$  at
$\log$\,\mst$\sim  11.8$.   These assumptions allow us to
parametrize  all toy-model properties (e.g.  the central DM
fraction) as a function of one   single quantity  (e.g. \Re, \mst,
or  $M_{dyn}$).
In Fig.~\ref{fig:fig_simulations},  we compare the  observed
correlations of  DM fraction  (upper panels) and density (lower
panels)  with \Re,  \mst, $M_{dyn}$ (from left to right in the
Figure),  to the toy-models' predictions,  where solid  and dashed
green  curves correspond  to  AC-NFW  and  NFW models,
respectively. Despite some differences in the trends, we find
qualitative agreement between the SIS-based observational results
and the toy-models, as the latter occupy  a region similar to that
of the data  in each diagram. On the other hand, adopting an
extreme \cMtoL\ model would produce disagreement between data and
theory, which predicts more DM on scales of $\sim$~10 kpc and
beyond.

\begin{figure}
\psfig{file=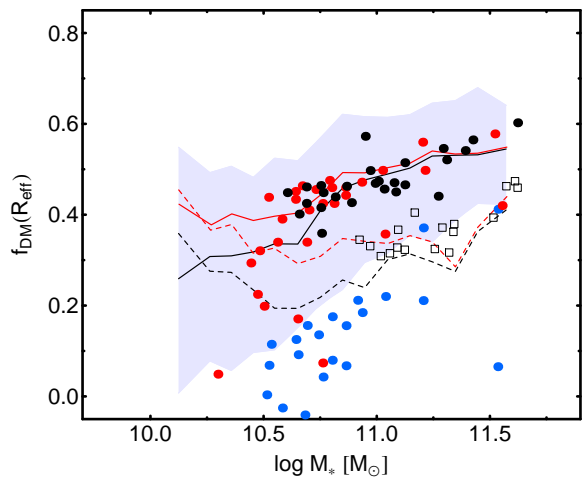,
  width=0.45\textwidth}\caption{Comparison of the \fdm--\mst\ relation
  to  the  predictions of  cosmological  simulations.  The black and  red
  curves,  as well  as  the grey  shaded  region, are  the  same as  in
  previous Figures.   Red and blue dots correspond  to simulated brightest
  cluster galaxies from~\citet{RS09},   for   contracted   and   uncontracted   models,
  respectively.    Black dots and  open squares   are  the   results  of
  hydrodynamical  simulations of~\citet{Onorbe+07}, adopting  high and
  low gas-to-stars conversion efficiency, respectively.}
\label{fig:fig_vs_Mstar_sims}
\end{figure}

In more detail, uncontracted (i.e.  NFW) models predict on average
smaller \fdm\ than contracted models (NRT10; T+10).  Both sets of
toy-models  give similar \fdm\  and \rhoDM\  predictions at high
mass (\Re), while they deviate more from  each other at lower mass
(\Re), with differences up to $\Delta  \fdm \sim 0.2$ and $\Delta
\log \rhoDM \sim 0.5  \, \rm \Msun$~kpc$^{-3}$.

The AC model matches the observations well over a broad range of
galaxy sizes and masses, which is remarkable since there was no
fitting or fine-tuning involved.  The NFW model fares less well,
particularly for galaxies with small masses and sizes.  As
discussed in Section~\ref{sec:dmdens}, the apparent implication is
that the central DM profiles have density cusps that are steeper
than in an NFW profile, and are broadly consistent with
adiabatically-contracted NFW profiles (see also NRT10; T+10). Such
conclusions will depend on the modelling ingredients, such as the
IMF adopted, the \eSF\ vs \mst\ relation etc., which it is outside
the scope of this paper to examine in detail. Here we can only
remark that one way to fix these discrepancies is to allow the
\eSF\ to grow with mass in contrast to the \cite{CW09}
predictions.

In    Fig.    \ref{fig:fig_vs_Mstar_sims}    we    compare   our
\fdm--\mst\ relation with the  results of two sets of
cosmological simulations: (1) simulated brightest cluster galaxies
from \citet{RS09}, using uncontracted and  contracted  DM models
(blue  and red  dots, respectively);    and  (2) hydrodynamical
simulations from \citet[hereafter ODS07]{Onorbe+07}, for the cases
where the authors adopted either a high  (black  dots) or low
(black empty squares) gas-to-stars conversion efficiency.

We find that contracted models   are  in   excellent agreement
with  the observed \SIS-based \fdm\ values, while uncontracted
models produce results closer to \cMtoL\ \fdm\ estimates, which is
qualitatively consistent with our toy-model results. The \SIS\
\fdm--\mst\  relation is also remarkably consistent with the
simulations  of~ODS07,  in  the case  of  high  conversion
efficiency. On the  other hand, \cMtoL\ \fdm's are  more
consistent to the predictions  of models  with low conversion
efficiency.  This is due to the fact that the low efficiency
simulations of~ODS07 are those producing lower \fdm's. In fact, as
noticed  by  the authors  (see  their Sec.~6.2), adopting a lower
conversion efficiency yields lower \Re\ values, which enclose  a
smaller portion of a galaxy DM  halo, implying lower \fdm\ values.
Finally,    one    should    notice    that   whilst    our
average \fdm--\mst\  relation is  fully  consistent with  that  of
ODS07,  the measured scatter  around it is  significantly larger
than that  of the simulations.  This might be due  to measurement
errors, as well as the effect  of some other  parameter which  is
not  taken into  account by simulations  (e.g.   galaxy   stellar
populations).  Moreover,
  consistently with the results in Fig.  \ref{fig:fig_simulations}, it
  is encouraging that our \Re-\mst\ relation is broadly consistent with
  the relation predicted by ODS07 for high efficiency models.

\section{Conclusions}\label{sec:conclusions}

We have discussed the central  DM content of massive, nearby ($z<0.1$)
ETGs,  using  high quality  data  from  the  SPIDER survey,  including
stellar masses  estimated from  optical- to NIR-photometry.  Using the
Jeans equations, we model each  galaxy to estimate its dynamical mass,
\Mdyn, within 1~\Re\ in the $K$-band.  We have adopted a \SIS\ and, as
a  comparison,  a non-homologous  \cMtoL\  model  for  the total  mass
profile, in order to  investigate the systematics induced by different
assumptions for the mass profile.  The recovered DM fraction, \fdm, or
equivalently the ratio $\Mdyn/\mst$ and the average densities, \rhoDM,
have  been  analyzed  as  a  function  of  structural  parameters  and
different  mass proxies,  comparing results  with predictions  of both
toy-models  and cosmological  simulations.  The  main results  are the
following:
\begin{itemize}
\item For a Chabrier IMF we  find that only small fractions ($\sim 12$
  and $24\%$ for the \SIS\  and \cMtoL, respectively) of galaxies have
  negative DM fractions, while the fractions increase to $\sim 55$ and
  $78\%$  when  a Salpeter  IMF  is  adopted.   Thus, under  the
    assumptions that the  total mass profiles are well  described by a
    \SIS\ or a \cMtoL\ and the  IMF is the same for all galaxies, our
  data favour  a Chabrier (or a \citealt{Kroupa01})  IMF, in agreement
  with,  e.g.,  \citealt{Cappellari06}.    On  the  other  hand,
    relaxing the assumption  of a universal IMF, for  the most massive
    galaxies in our  sample (with mass comparable to  the most massive
    early-type gravitational  lenses in  the SLACS survey)  a Salpeter
    IMF   is    consistent   with   observations   (\citealt{Treu+10};
    \citealt{Auger+10a}).  We remark that, at low mass, a Chabrier IMF
    is also preferred for late-type galaxies (\citealt{Sonnenfeld+11};
    \citealt{Brewer+12}), and  a Salpeter IMF  is ruled out  for ETGs,
    compared  to  a  Chabrier   IMF,  based  on  lensing  and  stellar
    population data~\citep{FSB:08, FER:10}.
\item  DM fractions  and densities  are,  on average,  higher when  an
  \SIS\   is   adopted,  when   compared   with   the  results   using
  \cMtoL\ profile. This result is also confirmed by the determinations
  of the  virial coefficient in Eq.   (\ref{eq:M_virial}), $K$, where,
  on average, $K_{\rm SIS} > K_{\rm M/L}$.
\item  The  DM fraction  is  a  steeply  increasing function  of  \Re,
  S\'ersic  index,  $\sigma$,  \mst\  and \Mdyn,  confirming  previous
  findings      (\citealt{Padmanabhan04};      \citealt{Cappellari06};
  \citealt{HB09};  T+09; \citealt{SLACSX};  \citealt{CT10}).  Galaxies
  with  denser  stellar  cores  have  also lower  \fdm\  values.   The
  assumption of a non-homologous \cMtoL\ profile yields weaker trends.
  The mass  dependencies may provide  important clues to the  FP tilt.
  Assuming that central DM trends can be combined with the virial
    theorem equation,  the \SIS\ results  point to a dominant  role of
    DM. If a  non homologous \cMtoL\ profile is  assumed, then almost
  all the tilt  would be due to a non-homology.  We have also verified
  that these  results are  qualitatively consistent with  our previous
  determinations (T+09 and NRT10).
\item  Similarly,  the central  average  DM  densities are  decreasing
  functions of  \Re, $n$ and masses.  Steeper trends are  found when a
  \cMtoL\   profile  is  adopted.   In  particular,   $\rhoDM  \propto
  \Re^{-1.8}$ and $\Re^{-2.2}$ for the \SIS\ and \cMtoL\ respectively,
  due to the steeper DM density profile for \cMtoL\ at \Re.
\item We have compared  our results with predictions from $\Lambda$CDM
  toy-models, finding good qualitative agreement. As in NRT10, we have
  found that when a Chabrier  (or Kroupa) IMF is adopted, a contracted
  NFW is  preferred, while a  bottom-heavy IMF like Salpeter  would be
  more consistent with an  uncontracted NFW (see, e.g. \citealt{CT10};
  NRT10; \citealt{Treu+10}).
\item  The  role  of  the  environment  has  been  investigated  after
  classifying  the galaxies in  the field  and in  groups, and  in the
  latter,  centrals have  also been  examined explicitly.  We  find no
  difference in the estimated  \fdm\ between field and group galaxies,
  and only slight differences  between centrals and non-centrals. This
  result  is possibly  due to  an increased  role of  mergers  for the
  central galaxies.
  \end{itemize}

We have finally shown that a very tight relationship similar to
the FP exists  between DM fraction,  velocity dispersion  and
size.   We have determined the best fitted  parameters in Eq.
(\ref{eq:DM_plane}) and have inferred that DM might be the
dominant driver of FP tilt, at
  least when a \SIS\ is adopted, and under the assumption that central
  DM trends can be combined  with the virial theorem expectation.  In
a forthcoming  paper we  will discuss the  results introduced  here in
terms of the stellar population parameters derived from the fitting of
spectra,   by  means   of  different   synthetic   prescriptions.   In
particular, within the general framework introduced in NRT10 and T+10,
we will  investigate correlations with  stellar populations parameters
such  as  age,   metallicity,  and  alpha-enhancement.   With  further
comparisons  to  simulations,  these   correlations  may  be  used  to
constrain  the evolution of  stars and  DM in  local galaxies,  and to
point the way to similar analyses at higher redshifts.

\section*{Acknowledgments}
We  thank the anonymous referee  for his/her kind  report and the
  help to improve the paper. We thank Jose O$\rm\tilde{n}$orbe to have
  provided us with data from  his simulations and for the discussions.
  Finally,  we also  thank Tommaso  Treu, Matthew  Auger,  Gary Mamon,
  Ignacio Ferreras,  Alister Graham, Cecile Faure and  Alexander V. Tutukov
  for their interest  in our work and the  feedback they provided us.
CT was  supported by  the Swiss National  Science Foundation.  AJR was
supported  by  National  Science  Foundation  Grants  AST-0808099  and
AST-0909237.

\appendix

\section{Structural parameters and magnitude}\label{sec:str_params}
\begin{figure*}
\psfig{file=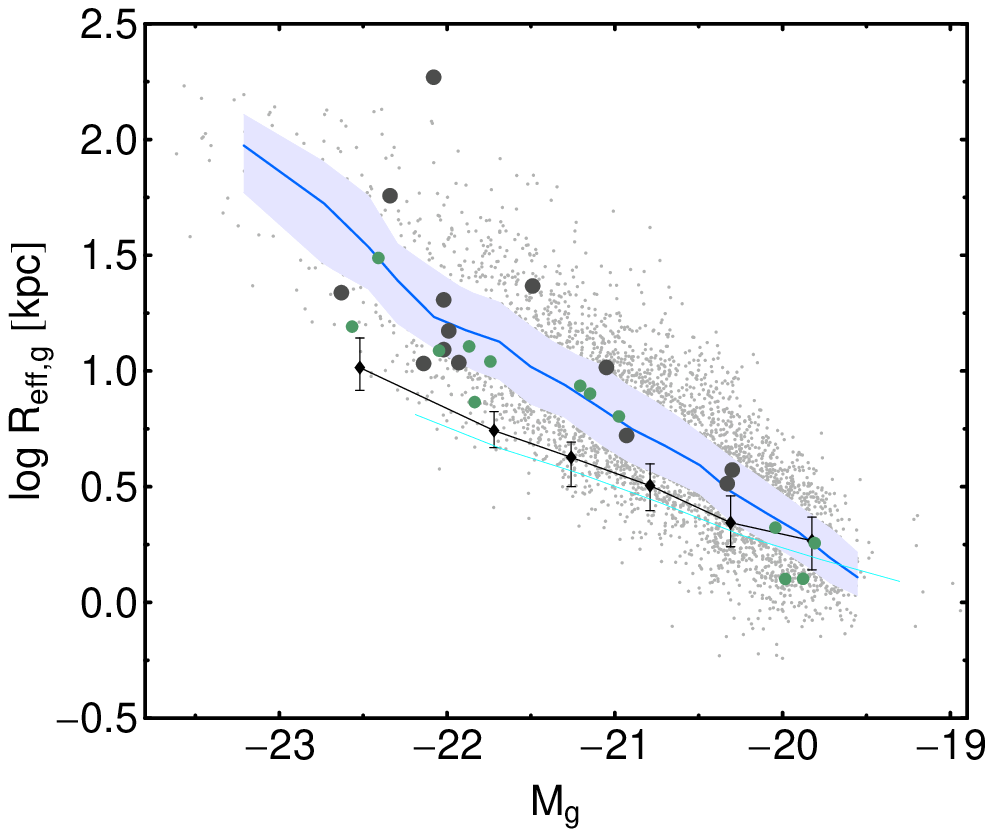,
width=0.43\textwidth}\psfig{file=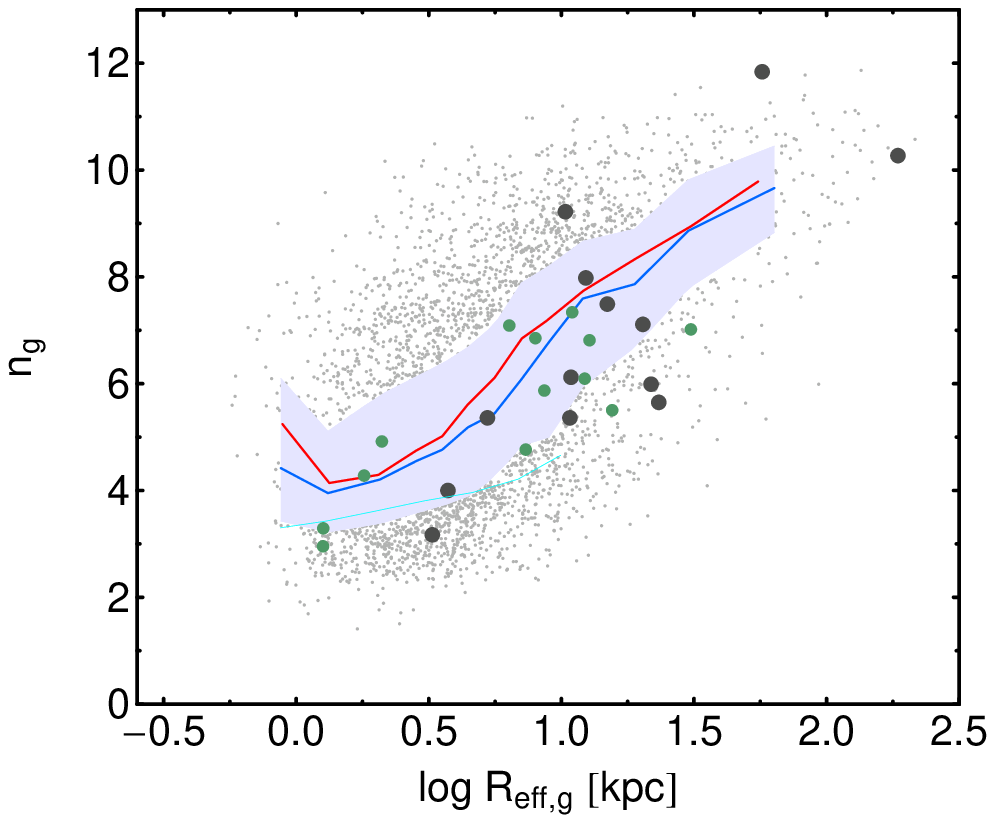,
width=0.43\textwidth}\caption{$\Re -  M_{g}$ (left) and $n_{g}  -
\Re$ (right) correlations.  The small grey points are the
datapoints in our sample, while the blue lines and shaded regions
are the  medians and $25-75$ per cent scatter. The red line in the
right plot is for the $r$-band  data. Large grey points are for
bright galaxies in \citet{Kormendy+09}. Black points with error
bars are the $B$-band results from T+09 (using the sample in
\citet{PS96}), while the cyan lines show the results using
NYU-VAGC data (\citealt{Blanton05}), and the green points are the
datapoints for the  Virgo  sample in \citet{Ferrarese+06}.
Suitable conversion factors  are adopted to  convert $r$-,  $B$-
or $V$-band data into $g$-band, and  the data   have  been
converted   to  our adopted cosmology. \label{fig:fig1}}
\end{figure*}

Structural parameters of ETGs, such  as the effective radius, \Re,
are known to  be significantly affected by the method (e.g. 1D vs.
2D fitting  of  the  light  distribution)  and profile shape (e.g.
de Vaucouleurs vs. S\'ersic law) used to derive them~(e.g.
\citealt{Kelson:00}).  While such  differences do not affect the
FP relation  of  ETGs, because of  the correlated variation  of
\Re\ and the mean surface brightness therein, $\langle \!\mu\!
\rangle_e$, they can affect other correlations among structural
parameters and luminosity (mass), and hence the correlations with
DM fractions (densities).
For instance,  in Sec.~\ref{sec:DM_fraction}  we  have  shown
that the \fdm\ correlations presented in the present study are
somewhat steeper than those we previously derived in T+09, based
on a different sample of ETGs,  from~\citet[hereafter PS96]{PS96}.
We address this issue in Fig.~\ref{fig:fig1},  where  we  compare
our  $g$-band luminosity--size relation to that  of the PS96
$B$-band sample analyzed  by T+09. For the sake of completeness,
in the same plot we include our $g$-band S\'ersic index--size
relation,  and compare both  relations to the  $g$-band data
from~\citet{Ferrarese+06}    and~\citet{Kormendy+09}.    Notice
that the effective parameters of  the PS96 sample are based  on de
Voucouleurs (rather than S\'ersic)  galaxy profile fits, hence no
comparison to our S\'ersic  $n$--size  relation  is  possible. We
refer  the  reader  to previous papers of the series (Papers I,
II) for a detailed comparison of correlations among structural
parameters from the SPIDER survey and SDSS  pipeline.
The present  sample  of  ETGs  exhibits  a  steeper
luminosity--size relation  than that  of  the  T+09  sample. This
is explained  by the fact  that  fitting a  high-$n$  galaxy  with
a  de Vaucouleurs profile   gives  a  systematically   smaller
\Re\ value, hence flattening the  luminosity--size relation (e.g.
\citealt{GW08}). Indeed,  this explains why our \fdm--\Re\
relation is steeper than that of T+09 (see
Fig.~\ref{fig:fig_fDM_comparison}), as a smaller \Re\ encloses a
smaller portion of the DM halo (implying lower \fdm\ estimates).
On  the  other hand, the correlations among structural parameters
for the present sample are  in better agreement with those
of~\citet{Ferrarese+06} and~\citet{Kormendy+09}.
We remark that since significant
differences in  the \fdm\  correlations exist when  analyzing
different  samples of  ETGs (i.e.   different sets of structural
parameters),  it is  of  paramount  importance to  compare
different samples of galaxies  (e.g.  different environments)
based on a homogeneous set of measurements, as the one used here.

\section{Dynamical mass procedure}\label{sec:Jeans_procedure}
Our basic approach to estimating the  DM content of ETGs is the
same as   in   T+09.    For   each   galaxy,  we   first
de-project   the corresponding~\citet{Sersic68}   light   profile,
assuming   spherical symmetry. This  gives the de-projected luminosity profile,
$j_*(r)$, where  $r$ is the spatial distance to the centre of the
galaxy. Using a S\'ersic model here has the main advantage of taking
into account the non-homology of the stellar  matter distribution
of  ETGs (see, e.g.,~\citealt{Caon93} and~\citealt{PS97}).  We
also assume some  simplified functional form for   the   dynamical
mass   profile,   $M(r)$,   i.e.    either   a \cMtoL\ profile,
$M(r) = \Upsilon_0 \,  L(r)$, or a  SIS model, where $M(r)\propto
\sigma_{\rm SIS}^{2} r$ (see Sec.~\ref{sec:sample}).
Assuming  spherical  symmetry and  no  rotation,  we  write the  Jeans
equation as:
\begin{equation}
{{\rm d}(j_* \sigma_r^2) \over {\rm d}r} + 2\,{\beta(r) \over r}
\,j_* \sigma_r^2 = - j_*(r)\, \frac{GM(r)}{r^2} \ ,
\label{eq:jeans}
\end{equation}
where $\beta = 1 - \sigma_t^2/\sigma_r^2$ is the anisotropy. Under
the hypothesis  of  isotropy (i.e.,  $\beta =  0$),
Eq.~(\ref{eq:jeans}) simplifies to
\begin{equation}
\sigma_r^2 (r) = \frac{1}{j_*(r)} \int_r^\infty j_* \frac{GM}{s^2}
{\rm d}s \ . \label{eq:iso}
\end{equation}
In order to  fit the given mass to the data,  we project this equation
in 2D, obtaining the line-of-sight velocity dispersion:
\begin{equation}
\sigma_{\rm los}^2 (R) = \frac{2}{I(R)}\,\int_R^\infty \frac{j_*
\sigma_r^2 \,r\,{\rm d}r}{\sqrt{r^2\!-\!R^2}} , \label{eq:siglos}
\end{equation}
where
\begin{equation}
I(R) = 2\,\int_R^\infty \frac{j_*\,r}{\sqrt{r^2\!-\!R^2}}{\rm d}r
\label{eq:IR}
\end{equation}
is   the   projected  surface   brightness   profile.   We   integrate
$\sigma_{\rm los}$ within a fixed aperture $R_{\rm Ap}$ (i.e. the SDSS
fibre   aperture)  to   obtain  the   aperture   velocity  dispersion,
$\sigma_{\rm Ap}$:
\begin{equation}
\sigma_{\rm Ap}^2 (R_{\rm Ap}) = \frac{1}{L(R_{\rm
Ap})}\int_0^{R_{\rm Ap}} 2\pi\,S\,I(S)\,\sigma_{\rm
los}^2(S)\,{\rm d}S \ , \label{eq:sigap}
\end{equation}
where  $L(R) =  \int_0^R 2\pi  S I(S)\,  {\rm d}S$  is  the
luminosity within the projected radius $R$. To avoid lengthy
calculations, we have adopted the compact formulae for
$\sigma_{\rm Ap}$ calculated in \cite{ML05} and
\cite{ML06_erratum} (see \cite{ML05b} for the anisotropic case).

Finally,  we   fit  the  model  $\sigma_{\rm  Ap}$   to  the
observed $\sigma_{spec}$,   by    varying   the   free parameters
in   Eq. (\ref{eq:iso})~  (i.e.   either $\sigma_{\rm SIS}$  or
$\Upsilon_0$) until the  desired matching is achieved. The
resulting best-fit mass profile provides the dynamical (spherical)
mass-to-light ratio within \Re,  \Ydyn\ (which coincides  with
$\Upsilon_0$  in  the case  of  a \cMtoL\ model), and mass \Mdyn.
Finally, as we have shown in Sec. \ref{sec:K} we notice that a non
null radial anisotropy (i.e. $\beta
> 0$) decreases the \Mdyn, lowering the DM fractions (see text for further details).

\end{document}